\documentclass[useAMS,usenatbib]{mn2e}
\usepackage{aas_macros}
\usepackage{natbib}
\usepackage{amssymb}
\usepackage{threeparttable}  
\usepackage{booktabs}
\usepackage{multirow}
\usepackage{bm}
\usepackage{amsfonts}
\usepackage{graphicx}
\usepackage{epstopdf}
\usepackage{longtable}
\usepackage{setspace}
\usepackage{ulem}
\bibpunct[, ]{(}{)}{,}{a}{}{,}

\newcommand{\Eexc}{$E_{\rm exc}$}
\newcommand{\Teff}{$T_{\rm eff}$}  % Command for Teff, text mode
\newcommand{\kms}{km\,s$^{-1}$}
\newcommand{\scm}{s$^{-1}$\,cm$^{-3}$}
\def\ione{\,{\sc i}}
\def\ii{\,{\sc ii}}
\def\iii{\,{\sc iii}}

\title[The interpretation of C\ione\ emission lines]
  {NLTE carbon abundance determination in selected A- and B-type stars and the interpretation of C\ione\ emission lines }
\author[S.~A.~Alexeeva, T.~A.~Ryabchikova and L.~I.~Mashonkina]
  {S.~A.~Alexeeva,$^1$ T.~A.~Ryabchikova,$^1$ $\&$ L.~I.~Mashonkina$^1$ 
  \newauthor % starts a new line in the
  $^1$Institute of Astronomy of the Russian Academy of Sciences \\
    48 Pyatnitskaya St. 119017, Moscow, Russia}
\date{Released 2014 Xxxxx XX}

\pagerange{\pageref{firstpage}--\pageref{lastpage}} \pubyear{2014}

\def\LaTeX{L\kern-.36em\raise.3ex\hbox{a}\kern-.15em
    T\kern-.1667em\lower.7ex\hbox{E}\kern-.125emX}

\begin{document}

\label{firstpage}

\maketitle

\begin{abstract}

We constructed a comprehensive model atom for C\ione\ -- C\ii\ using the most up-to-date atomic data available and evaluated the non-local thermodynamic
equilibrium (NLTE) line formation for C\ione\ and C\ii\ in classical 1D models representing the atmospheres of A and late B-type stars. Our NLTE calculations predict the emission that appears at effective temperature of 9250 to 10\,500~K depending on log~$g$ in the C\ione\ 8335, 9405\,\AA\ singlet lines and at \Teff~$>$~15\,000~K (log~$g$ = 4) in the 
C\ione\ 9061 -- 9111\,\AA\,, 9603 -- 9658\,\AA\, triplet lines. A prerequisite of the emission phenomenon is the overionization-recombination mechanism resulting in a depopulation of the lower levels of C\ione\ to a greater extent than the upper levels.
Extra depopulation of the lower levels of the transitions corresponding to the near-infrared
 lines, is caused by photon loss in the UV lines C\ione\ 2479, 1930, and 1657\,\AA. 
We analysed the lines of C\ione\ and C\ii\ in Vega, HD~73666, Sirius, 21~Peg, $\pi$~Cet,
HD~22136, and $\iota$ Her taking advantage of their observed high-resolution spectra.
The C\ione\ emission lines were detected in the four hottest stars,
 and they were well reproduced in our NLTE calculations.
For each star, the mean NLTE abundances from lines of the two ionization stages, C\ione\ and C\ii, including the C\ione\ emission lines, were found to be consistent. We show that the predicted C\ione\ emission phenomenon depends strongly on whether
accurate or approximate electron-impact excitation rates are applied.

\end{abstract}


\section{Introduction}

Rapid development of observational techniques over the last decade has resulted in dramatic improvement of the quality of spectral observations in astronomy. 
Thanks to Echelle spectrographs, we are able to cover wide spectral regions with one exposure, obtaining spectra with a high spectral resolution of up to $R$~=~$\lambda$/$\Delta\lambda$~=~120\,000 and high signal-to-noise ratio. Huge amounts of high quality spectra are collected in different archives with the open access for the astronomical community. 
High quality spectra require immediately an adequate improvement in theoretical methods of spectrum analysis, model atmospheres, line formation scenarios, etc.  

\citet{2009A&A...503..945F} studied three apparently normal A and B sharp-lined stars and found the C\ione\ emission lines at 8335, and 9406\,\AA\, in the hottest of them, $\pi$~Cet. 
The absorption C\ione\ lines in the 7111--7120\,\AA\, range are rather shallow and allow us to determine only an upper limit for the abundance. In local thermodynamic equilibrium (LTE) analysis, $\pi$~Cet reveals a disparity between
C\ione\ and C\ii, in line with \citet{1990ApJS...73...67R}. 
For the cooler star, 21~Peg, \citet{2009A&A...503..945F} could not obtain consistent abundances from different C\ione\ lines, namely the abundances obtained from C\ione\ 4932, 5052, 5380~\AA\, were found to be significantly smaller than those from other C\ione\ lines. 

\citet{2009A&A...503..945F} discussed several mechanisms of the C\ione\ emission in  $\pi$~Cet. Emission lines may form, if the star has a chromosphere, however, there is no evidence for chromospheric activity in $\pi$~Cet. This star was classified as a Herbig Ae/Be star due to small infrared (IR) excess \citep{1998A&A...331..211M}, suggesting the existence of a circum-stellar disc. This is supported by detection of an emission signature in H$_\alpha$ \citep{2009A&A...503..945F}.
However, half-widths of the C\ione\ emission lines do not differ from those of the C\ione\ absorption lines, giving evidence for their common origin in the star's atmosphere.
\citet{2009A&A...503..945F} proposed that the emission can be caused by 
the departures from LTE in the C\ione\ line formation. In the literature there are examples of successfully reproducing the observed emission lines by non-local thermodynamic equilibrium (NLTE) calculations, for example, Mg\ione\ 12~$\mu$m in the Sun \citep{Carlsson92}, Mn\ii\ 6122-6132\,\AA\ in the three late type B stars \citep{2001A&A...377L..27S}, C\ii\ 6151, 6462\,\AA\ in $\tau$~Sco (B0V) and C\ii\ 6462\,\AA\ in HR~1861 (B1V) \citep{2008A&A...481..199N}.     

There are few NLTE studies of C\ione/C\ii\ in the early A to late B stars.
\citet{1996A&A...312..966R} computed a negative and small absolute value of less than 0.05~dex NLTE abundance corrections for lines of C\ii\ in A-type stars.
\citet{2001AA...379..936P} obtained very small NLTE corrections for the C\ione\ and C\ii\ lines in the visible region using a Vega model atmosphere with an effective temperature \Teff\ = 9550~K.
 The NLTE carbon abundance analysis of 20 sharp-lined early B stars in the effective temperature range of 16000--33000~K, including two stars with \Teff\ $<$ 18000~K, was performed by \citet{2012A&A...539A.143N}.    
None of the carbon NLTE papers investigated the NLTE effects for the C\ione\ near-IR lines in the early A to late B stars.

This paper aims to understand mechanism(s) of the C\ione\ emission in B-type stars and to treat the method of accurate abundance determination from different lines of C\ione\ and C\ii\ based on NLTE line formation.
We construct a comprehensive model atom for C\ione\ -- C\ii\ using the most up-to-date atomic data available so far and analyse lines of C\ione\ and C\ii\ in high-resolution spectra of reference A and B-type stars with well-determined stellar parameters.

The paper is organized as follows. Section\,\ref{Sect:atom} describes an updated model atom of C\ione-C\ii and discusses departures from LTE for C\ione\ -- C\ii\ in the model atmospheres of A-B stars and mechanisms driving the C\ione\ emission lines. 
In Sect.\,\ref{Sect:Stars}, we analyse the C\ione\ near-IR emission lines observed in the four B-type stars and determine the C abundance of the selected A-B stars. We inspect the abundance differences between different lines of a common species and between two ionisation stages, C\ione\ and C\ii.
Our conclusions are summarized in Sect.\,\ref{Sect:Conclusions}.

\begin{figure*}
\begin{minipage}{160mm}
%\setcaptionmargin{5mm}
%\onelinecaptionsfalse
\includegraphics[scale=0.6]{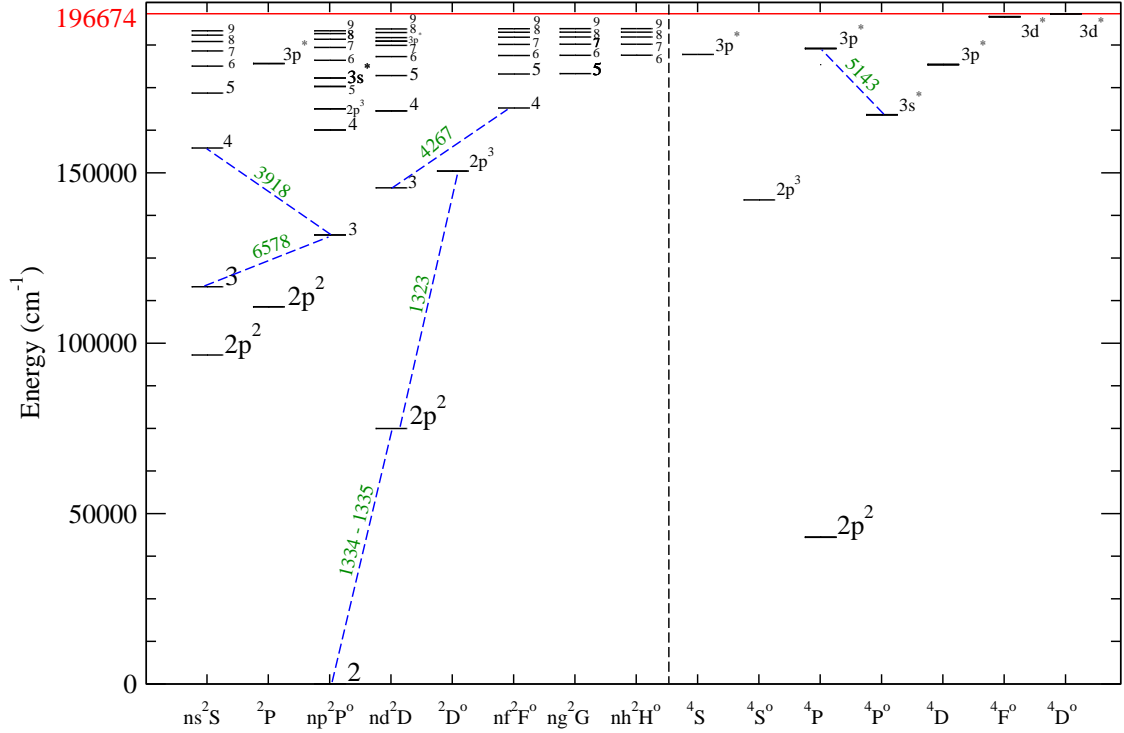}
%\captionstyle{normal}
\caption{The term diagram for singly-ionized carbon. The dashed lines indicate the seven transitions, where the investigated spectral lines arise. }
\label{Grot}
\end{minipage} 
\end{figure*}

\begin{figure*}
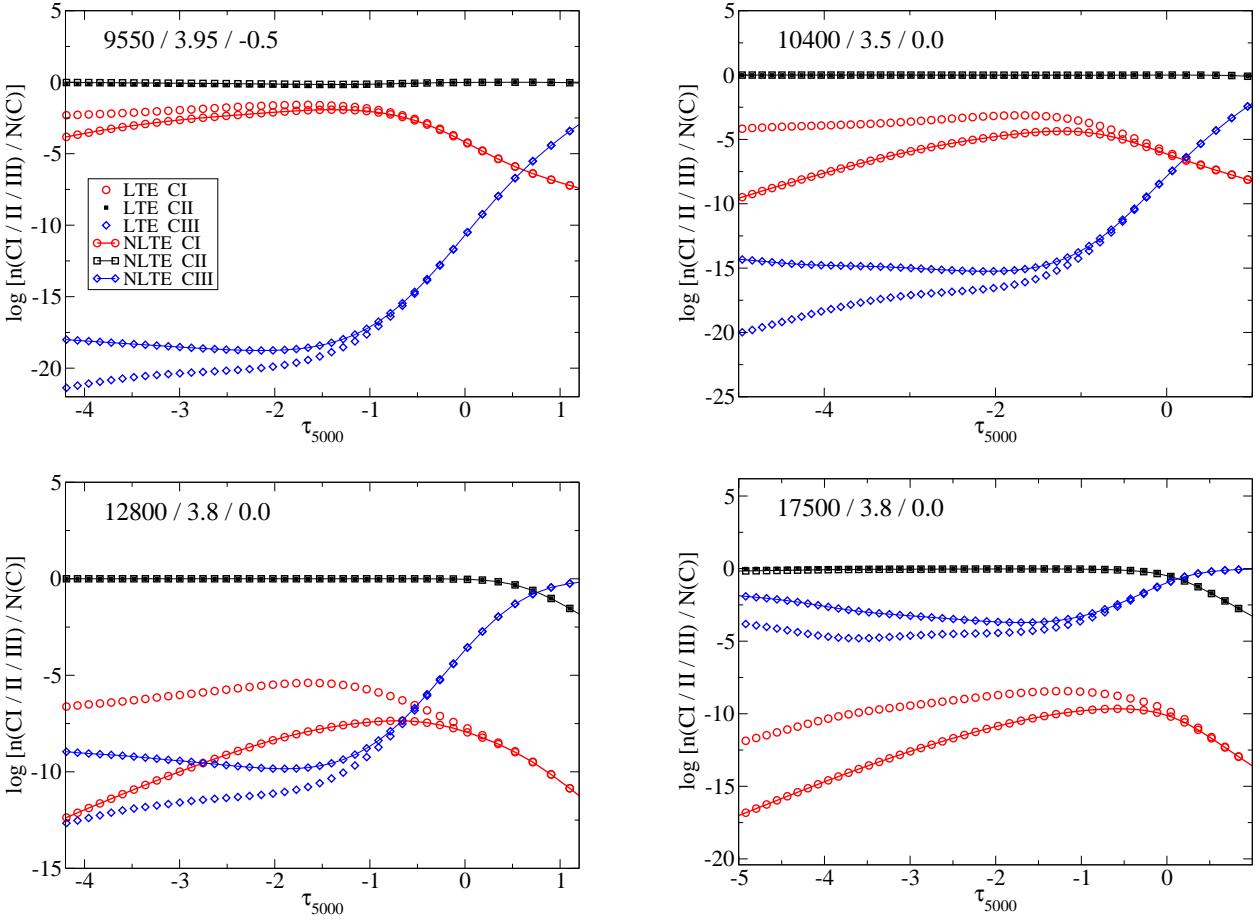

 \begin{minipage}{175mm}
\parbox{0.3\linewidth}{\includegraphics[scale=0.28]{Figure2.eps}\\
\centering}
\hspace{0.2\linewidth}
\parbox{0.3\linewidth}{\includegraphics[scale=0.28]{Figure3.eps}\\
\centering}
\hfill
\\[0ex]
\parbox{0.3\linewidth}{\includegraphics[scale=0.28]{Figure4.eps}\\
\centering}
\hspace{0.2\linewidth}
\parbox{0.3\linewidth}{\includegraphics[scale=0.28]{Figure5.eps}\\
\centering}
\hfill
\caption{NLTE and LTE fractions of C~I, C~II, and C~III in the model atmospheres of different effective temperature. }
\label{balance}
\end{minipage}
\end{figure*} 

\section{NLTE Line formation for C\ione-C\ii}\label{Sect:atom}

\subsection{Model atom and atomic data}\label{Sect:modelatom}

\citet{2015MNRAS.453.1619A} developed and described in detail the carbon model atom 
that represents the term structure of C\ione\ well and only approximately that of C\ii. Here, the model atom was updated by extending the C\ii\ term structure and including recent accurate collisional data for C\ione.

{\bf Energy levels of C\ii\,.} 
 Energy levels of C\ii\ belong to the doublet terms 
of the 2s$^2$~$nl$ ($n$ = 2 -- 10,  $l$ = 0 -- 2), 2s$^2$~$n$f ($n$ = 4 -- 10), 2s$^2$~$n$g ($n$ = 5 -- 10), 2s$^2$2p~$n$h ($n$ = 6 -- 10), 2s2p$^3$ , and 2s2p~$3l$ ($l$ = 0 -- 1) electronic configurations and the quartet terms of 2s2p$^2$, 2s2p$^3$, 2p$^3$, 2s2p~$3l$ ($l$ = 0 -- 3). We thus include explicitly the C\ii\ energy levels up to 0.01~eV below the ionization threshold.
Fine structure splitting was taken into account everywhere, up to $n$ = 5. 
All the C\ii\ states with $n$ = 11 to 13 have close together energies, therefore they were combined into superlevels.
%, and their populations must be in equilibrium to each other. 
%The levels of common parity were combined into the superlevels.  
Energy levels were adopted from the NIST\footnote{http://physics.nist.gov/PhysRefData/} database \citep{NIST08}.
The term diagram for C\ii\ is shown in Fig.\,\ref{Grot}.

{\bf Radiative data.} We included 2138 allowed bound-bound ($b-b$) transitions, whose transition probabilities were taken from the
NIST and VALD \citep{vald} databases, where available, and the Opacity Project 
(OP) database TOPbase\footnote{http$//$legacy.gsfc.nasa.gov$/$topbase} \citep{1993BICDS..42...39C, 1989JPhB...22..389L,  1993AAS...99..179H}. 
Their  Photo-ionization cross-sections  
for levels of C\ii\ with $n \le 10$, $l \le 3$ are from TOPbase, and we adopted the hydrogen-like cross-sections for the higher excitation levels.

{\bf Collisional data.} For all the transitions between levels of the C\ione\ 2p$^2$, 2p$^3$, 2p3$l$ ($l$ = 0, 1, 2),  and 2p4s electronic configurations we use the effective collision strengths computed by \citet{2013PhRvA..87a2704W} on the base 
of the $B$-spline $R$-matrix with pseudo states method. 
For the transitions involving the higher excitation levels of the 2p4p, 2p4d, and 2p5s electronic configurations we adopted cross-sections calculated by \citet{Reid1994} in the close-coupling approximation using the R-matrix method.
For the transitions connecting the 30 lowest levels in C\ii\ we use the effective collision strengths from \citet{2005AA...432..731W}. 
The remaining transitions in both C\ione\ and C\ii\ were treated using an approximation formula of \citet{1962ApJ...136..906V} for the allowed transitions and assuming the effective collision
strength $\Omega_{ij}$ = 1 for the forbidden transitions. 
Ionization by electronic collisions was everywhere treated through the \citet{1962PPS....79.1105S} classical path approximation.

\subsection{Method of calculations}

 We used the code \textsc{DETAIL} \citep{detail} based on the accelerated $\Lambda$-iteration method \citep{rh91}
for solving the radiative transfer and statistical equilibrium (SE) equations.
The opacity package was updated by \citet{2011JPhCS.328a2015P}.
The departure coefficients, $b_{\rm{i}}$ = $n_{\rm NLTE}$ / $n_{\rm LTE}$, were then used to calculate the synthetic NLTE line profiles via the \textsc{binmag3} code \citep{binmag3} and \textsc{synthV-NLTE} code \citep{2016MNRAS.456.1221R} . 
Here, $n_{\rm{NLTE}}$ and $n_{\rm{LTE}}$ are the SE and thermal (Saha--Boltzmann) number densities, respectively. 

We used plane-parallel (1D), chemically homogeneous model atmospheres calculated with the \textsc{LLmodels} code \citep{2004AA...428..993S}.
An exception is Sirius, for which we took the Kurucz's model\footnote{http://kurucz.harvard.edu/stars/SIRIUS/ap04t9850g43k0he05y.dat} computed with the atmospheric parameters close to those derived by \citet{1993AA...276..142H}. 

The carbon abundance was determined from line profile fitting. 
The uncertainty in fitting the observed profile is less than 0.02 dex for weak lines and 0.03 dex for strong lines.

\begin{figure*}
 \begin{minipage}{175mm}
%\begin{center}
\parbox{0.35\linewidth}{\includegraphics[scale=0.6]{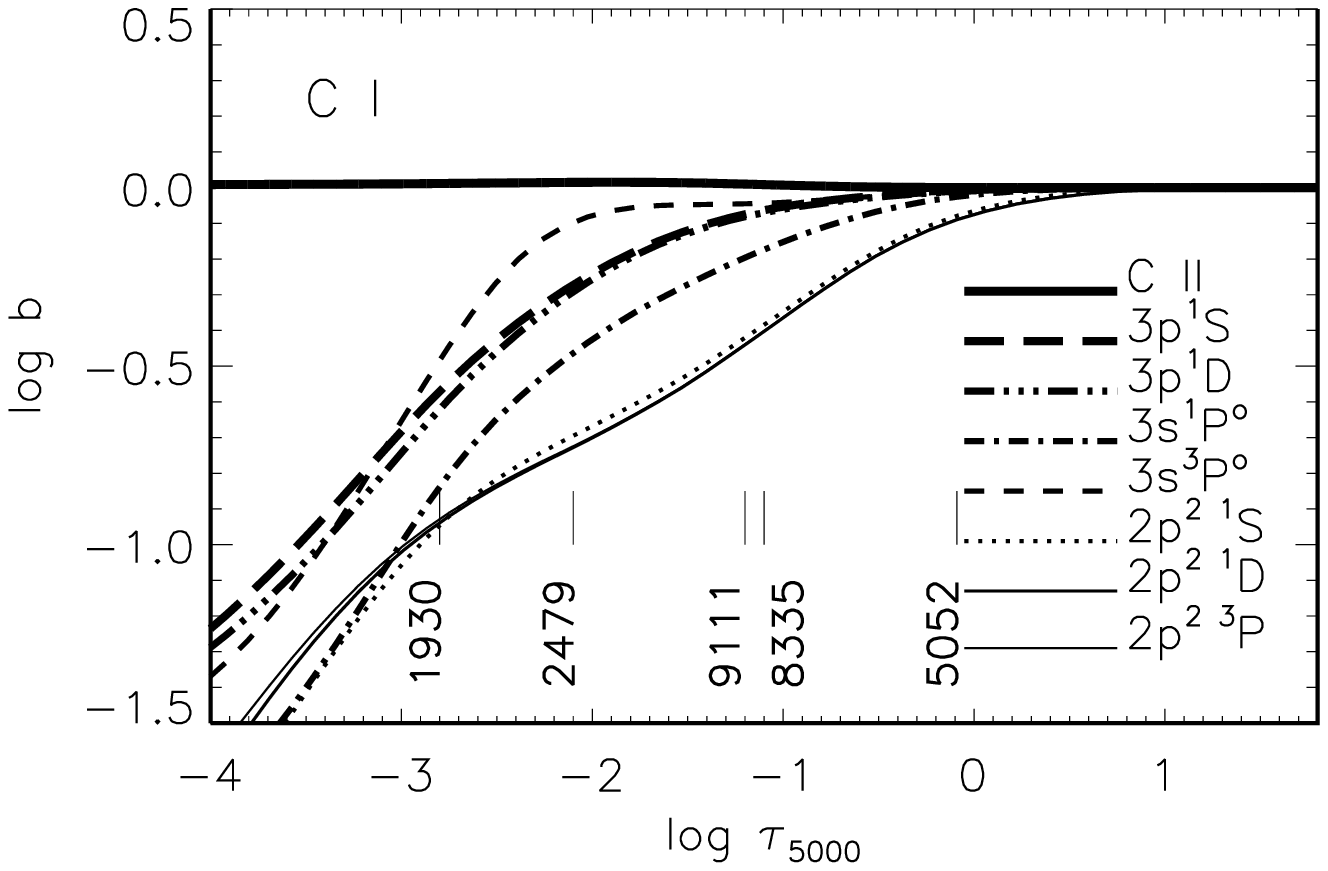}\\
\centering}
\hspace{0.1\linewidth}
\parbox{0.35\linewidth}{\includegraphics[scale=0.6]{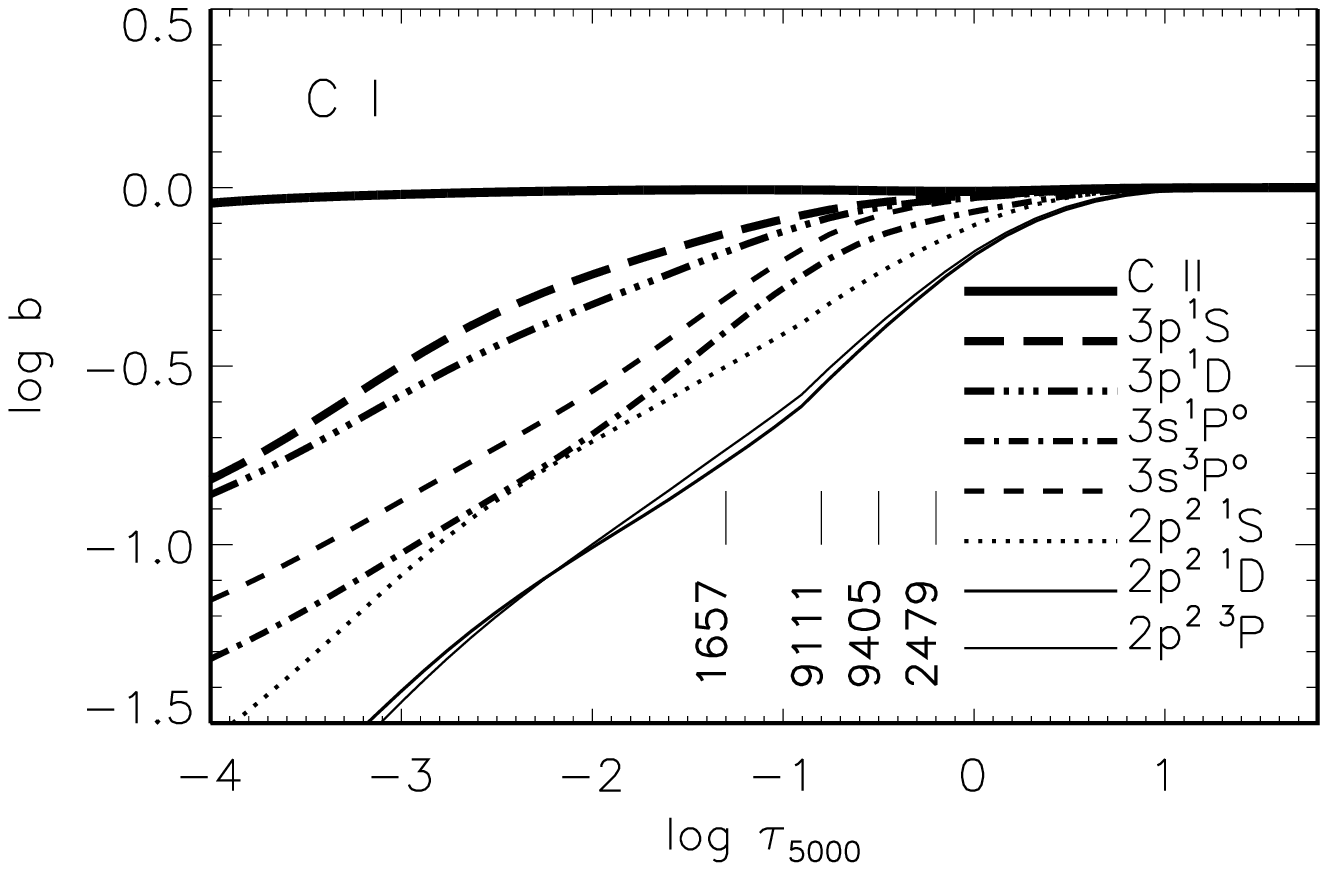}\\
\centering}
\hspace{0.00\linewidth}
\hfill
\\[0ex]
\parbox{0.35\linewidth}{\includegraphics[scale=0.6]{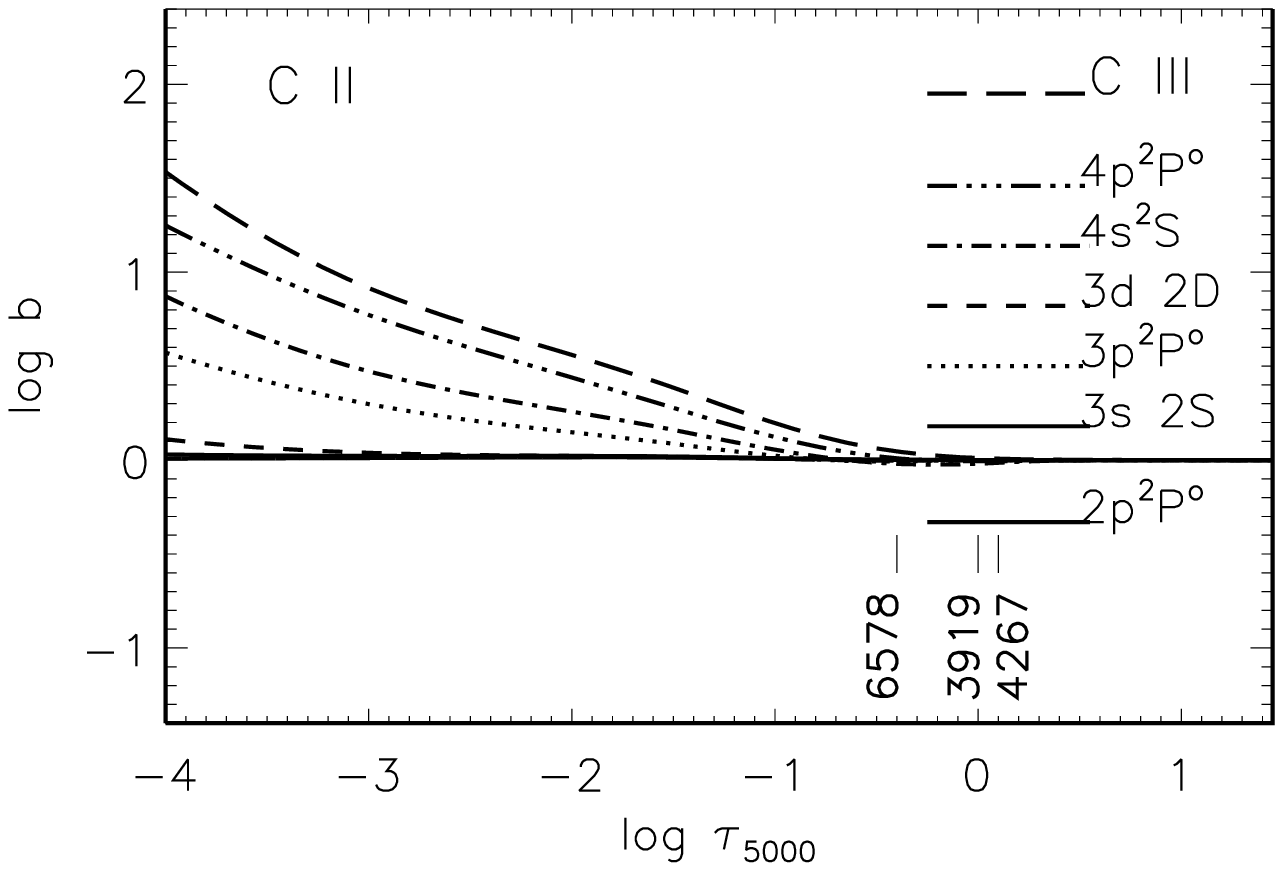}\\
\centering}
\hspace{0.1\linewidth}
\parbox{0.35\linewidth}{\includegraphics[scale=0.6]{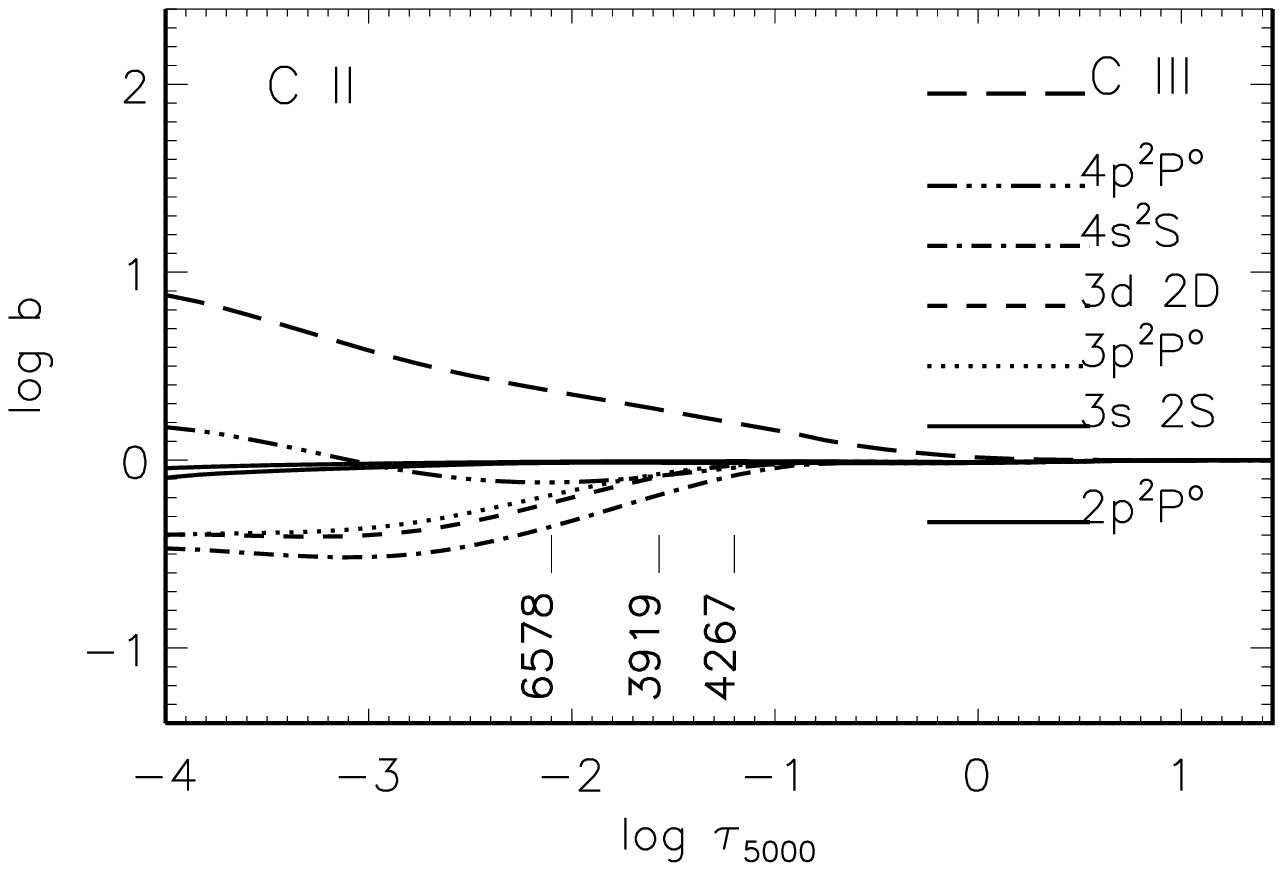}\\
\centering}
\hfill
\caption{Departure coefficients for selected levels of C\ione\ (top row) and C\ii\ plus the C\iii\ ground state (bottom row) as a function of $\log \tau_{5000}$ in the  model atmospheres 10400 / 3.5 / 0.0 (left-hand column) and 17500 / 3.8 / 0.0 (right-hand column). The wavelengths and vertical marks indicate the location of optical depth unity for the 
lines under consideration.
 }
\label{DC}
\end{minipage}
\end{figure*}

\subsection{SE of carbon}\label{Sect:departure}

In Fig.\,\ref{balance}, we show the LTE and NLTE fractions of C\ione, C\ii, and C\iii\ 
in the model atmospheres of different temperatures. 
In each model, C\ii\ dominates the total element abundance throughout the line--formation region. 
The fraction of C\ione\ everywhere does not exceed several thousandth parts, and in the line--formation layers, it is smaller in NLTE than in LTE due to ultraviolet (UV) overionization.

Fig.\,\ref{DC} displays a behaviour of the departure coefficients for selected levels in the model atmospheres with \Teff / log~g / [Fe/H] = 10400 / 3.5 / 0 and 17500 / 3.8 / 0.
Deep in the atmosphere, the departure coefficients 
are equal to unity because of a dominance of the collision processes in establishing the statistical equilibrium.
In the line--formation layers, outward log~$\tau_{5000}$ = 0.5, NLTE leads to depleted populations of the C\ione\ levels  because of UV overionization.

The deviations from LTE are small everywhere for the ground state of C\ii.
In the model 10400/3.5/0, outwards log~$\tau_{5000} = -0.5$, the C\ii\ high-excitation levels  
3p$^{2}$P$^{\circ}$, 4s$^{2}$S, and 4p$^{2}$P$^{\circ}$) are overpopulated via the UV radiative pumping transitions 2p$^{2}$ $^{2}$D -- 3p$^{2}$P$^{\circ}$ (C\ii\ 1760\,\AA), 3p$^{2}$P$^{\circ}$ -- 4s$^{2}$S (C\ii\ 3918\,\AA), and 
2p$^{2}$ $^{2}$D -- 4p$^{2}$P$^{\circ}$ (C\ii\ 1142\,\AA).
In the model 17500/3.8/0, these UV lines form in higher atmospheric layers, around log~$\tau_{5000} = -1.0$, resulting in a depopulation of the upper levels via spontaneous transitions.

The NLTE effects for a given spectral line can be understood from
analysis of the departure coefficients of the lower and upper levels, b$_l$ and b$_u$, at the line formation depths. The line can be strengthened or weakened compared with its LTE strength depending on the b$_l$ and b$_u$ values. Here, we consider two lines with similar excitation energy of the lower level, \Eexc, but different $gf$ values. Atomic data for the investigated lines are indicated in Table~\ref{tab1}.
In the model 10400/3.5/0, the C\ione\ 5052\,\AA\ line (3s$^{1}$P$^{\circ}$ -- 4p$^{1}$D) is weak and it forms in the layers around $\log\tau_{5000} = 0$, where b$_{l} < 1$ and b$_{l} < {\rm b}_{u}$. NLTE leads to weakening this line and positive NLTE abundance correction of $\Delta_{\rm NLTE}$ = log$\epsilon_{\rm NLTE}$ -- log$\epsilon_{\rm LTE}$ = +0.45 dex. In contrast, C\ione\ 9111\,\AA\ (3s$^{3}$P$^{\circ}$ -- 3p$^{3}$P) is strong and its core forms around $\log\tau_{5000} = -1.2$, where b$_{l} > {\rm b}_{u}$ and the line source function drops below the $Planck$ function resulting in strengthened line and negative NLTE abundance correction of $\Delta_{\rm NLTE} = -0.13$.

\subsection{Mechanisms driving the C\ione\ emission}

Our calculations predict that the C\ione\ near-IR lines at 8335\,\AA\ (3s$^{1}$P$^{\circ}$ -- 3p$^{1}$S), 9405\,\AA\ (3s$^{1}$P$^{\circ}$ -- 3p$^{1}$D), 9061 -- 9111\,\AA\ (3s$^{3}$P$^{\circ}$ -- 3p$^{3}$P), and 9603 -- 9658\,\AA\ (3s$^{3}$P$^{\circ}$ -- 3p$^{3}$S) may appear as emission lines depending on the atmospheric parameters.

\begin{figure*}
 \begin{minipage}{190mm}
\begin{center}
\parbox{0.38\linewidth}{\includegraphics[scale=0.25]{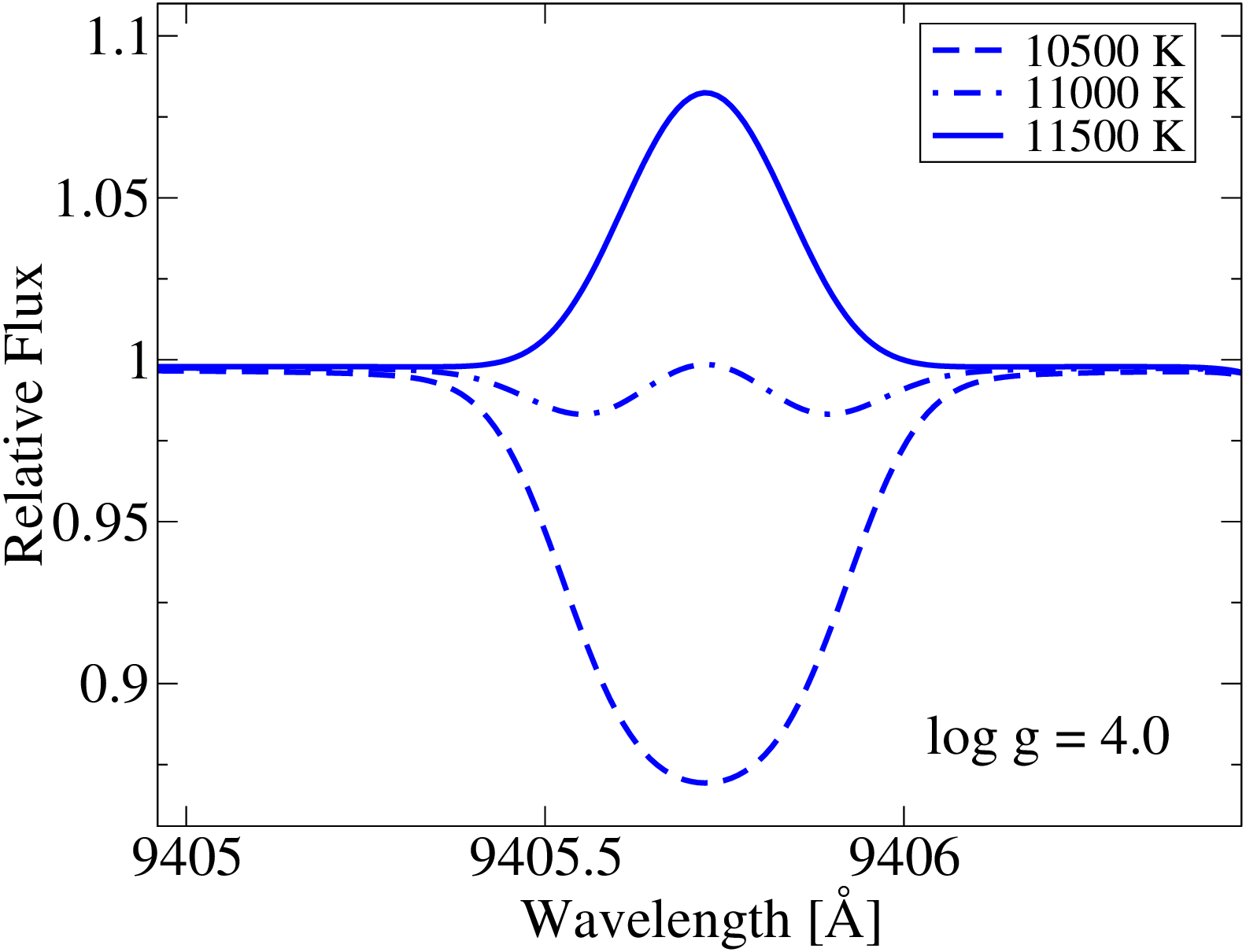}\\
\centering}
\parbox{0.38\linewidth}{\includegraphics[scale=0.25]{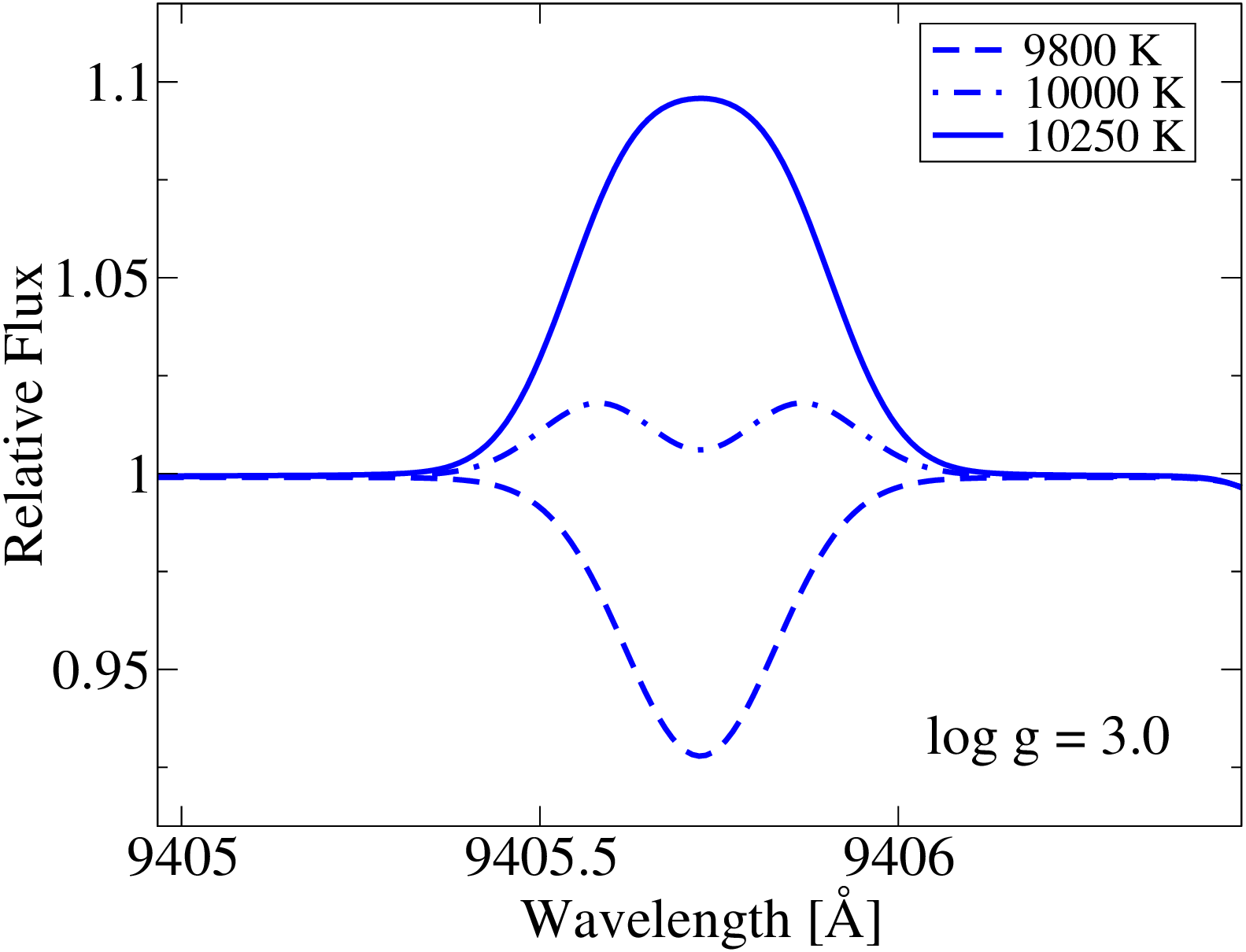}\\
\centering}
\hspace{1\linewidth}
\hfill
\\[0ex]
\parbox{0.38\linewidth}{\includegraphics[scale=0.25]{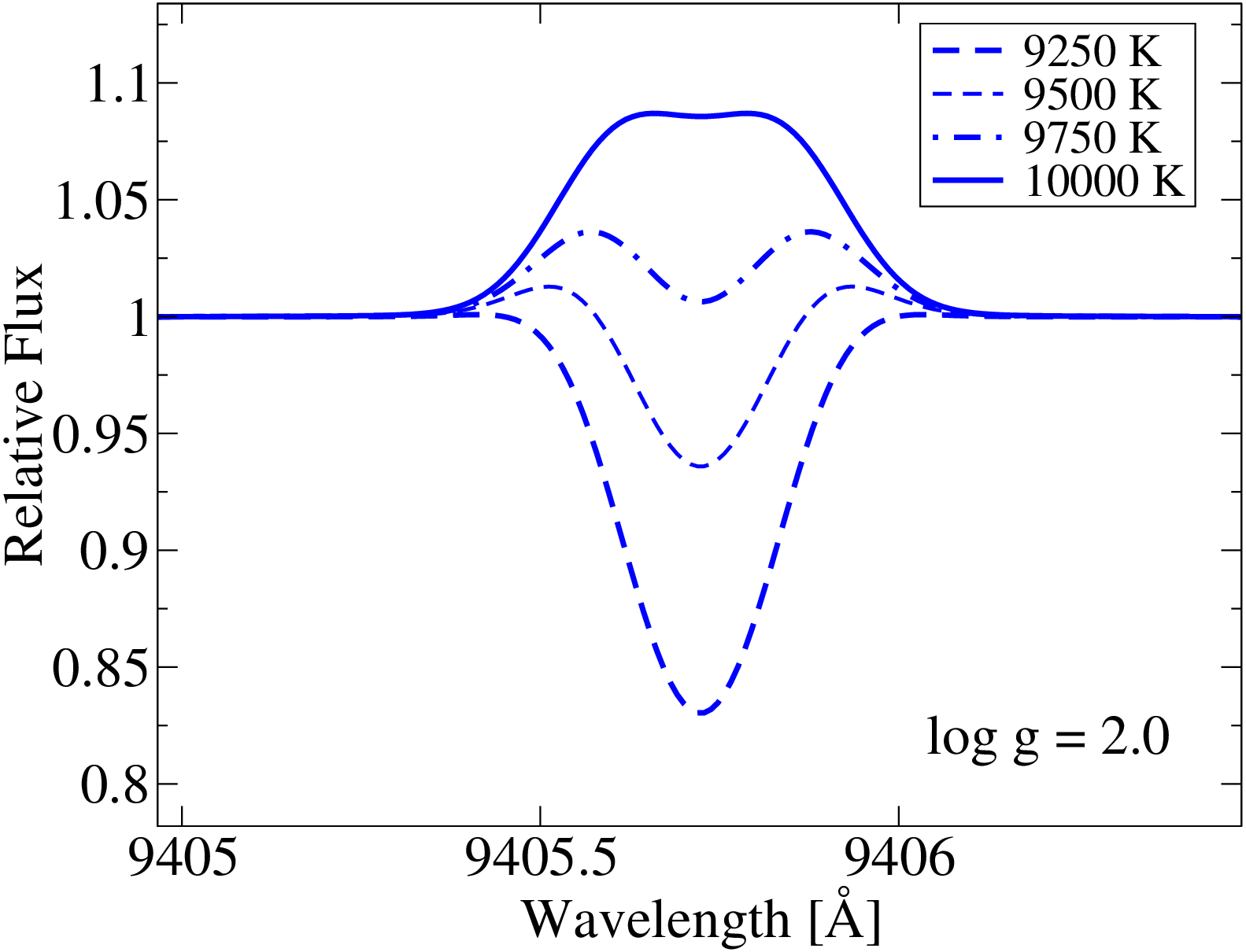}\\
\centering}
\parbox{0.38\linewidth}{\includegraphics[scale=0.25]{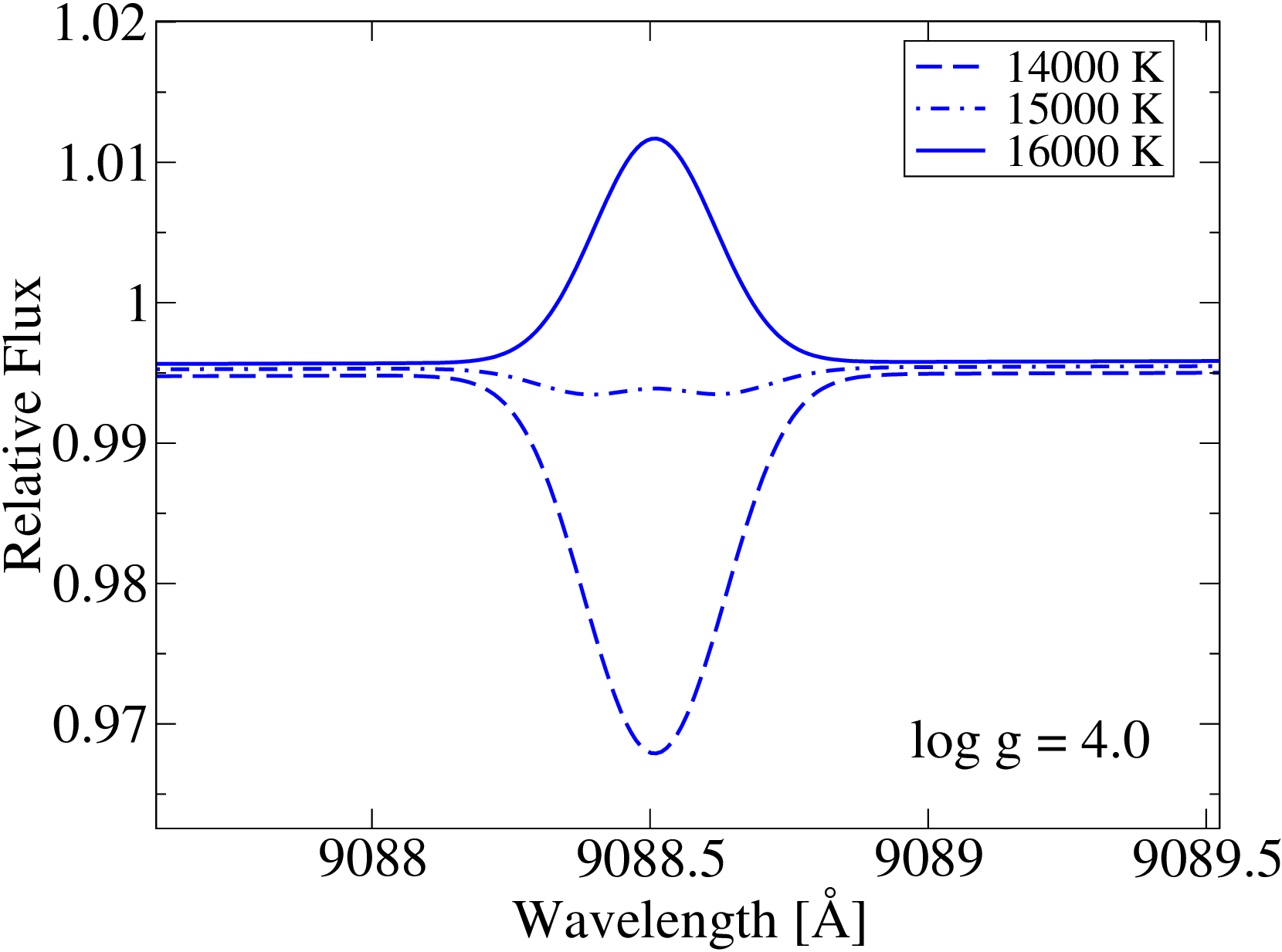}\\
\centering}
\hspace{1\linewidth}
\hfill
\\[0ex]
\parbox{0.38\linewidth}{\includegraphics[scale=0.25]{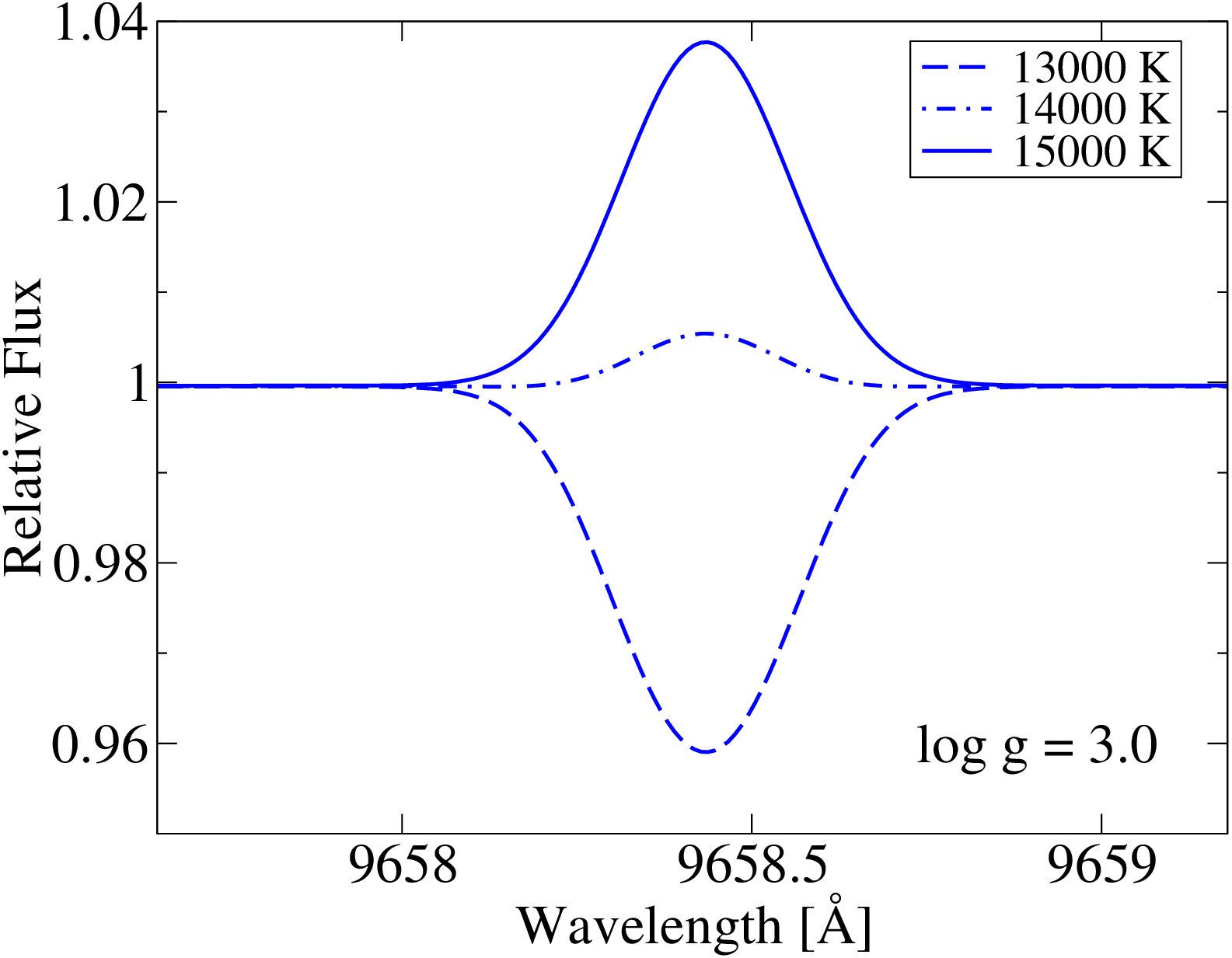}\\
\centering}
%\hspace{0.00\linewidth}
\parbox{0.38\linewidth}{\includegraphics[scale=0.25]{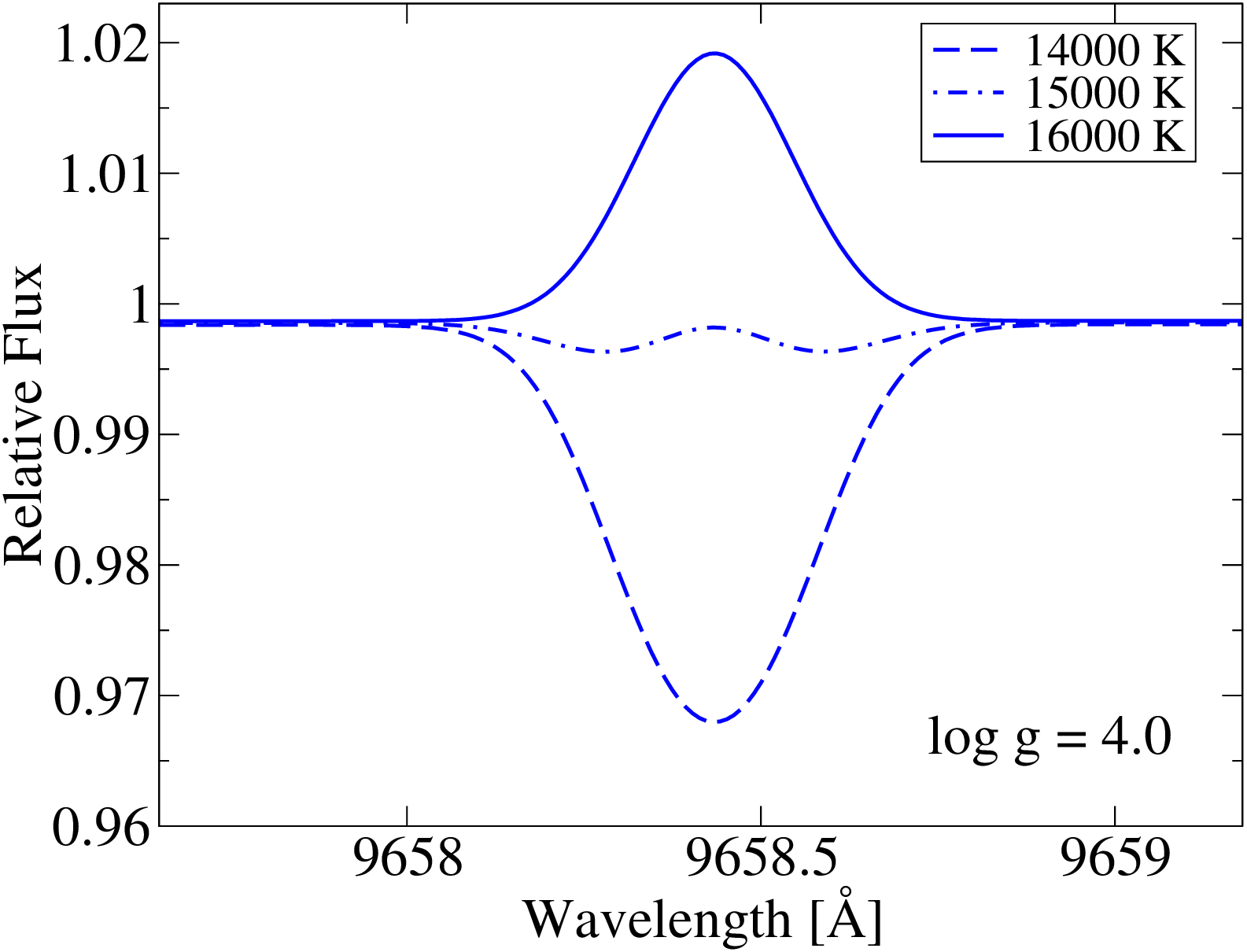}\\
\centering}
\hspace{1\linewidth}
\hfill
\\[0ex]
\caption{Change of the line profile with a variation in \Teff\ and log~$g$ for C\ione\ 9405\,\AA, 9658\,\AA, and 9088\,\AA.
Everywhere, [C/Fe] = 0, $V$sin$i$ = 0~\kms,  and $\xi_t$ = 1~\kms. The theoretical spectra are convolved with an instrumental profile of R = 65\,000.}
\label{param34}
\end{center}
\end{minipage}
\end{figure*} 

\begin{figure}
\begin{center}
\includegraphics[scale=0.3]{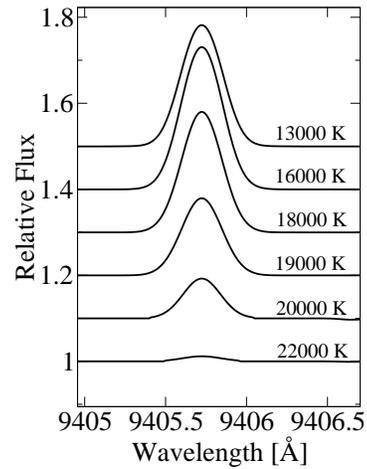}
\caption{Changes in the C\ione\ 9405\,\AA\ line profile with temperature increasing. Everywhere, log~$g$ = 3, [C/Fe] = 0, $V$sin$i$ = 0~\kms, and $\xi_t$ = 1~\kms. The theoretical spectra are convolved with an instrumental profile of R = 65\,000. }
\label{evol}
\end{center}
\end{figure}

The C\ione\ 8335 and 9405\,\AA\ singlet lines reveal an emission at the lower effective temperature compared with the triplet lines. The lower log~$g$, the lower \Teff\ is, at which the emission profile is developed for given line.
Fig.~\ref{param34} shows how a variation in \Teff\ and log~$g$ affects theoretical profiles of C\ione\ 9405, 9658, and 9088\,\AA\ and at which atmospheric parameters the line absorption changes to emission. For example, the C\ione\ 9405\,\AA\ emission line appears at 9250~K $<$ \Teff\ $<$ 9500~K in the log~$g$ = 2 models and at \Teff\ $>$ 10\,500~K when log~$g$ = 4. In contrast, the triplet line at 9658\,\AA\ comes into emission at \Teff\ $>$ 15\,000~K in the log~$g$ = 4 models. Having appeared at specific \Teff/log~$g$, the emission is strengthened towards higher temperature, reaches its maximum and gradually disappears. For example, in the models with log~$g$ = 3, the strongest emission in C\ione\ 9405\,\AA\ is predicted for \Teff\ = 16\,000~K, and it almost disappears at \Teff\ = 22\,000~K (Fig.~\ref{evol}). 

\begin{figure*}
\includegraphics[scale=0.6]{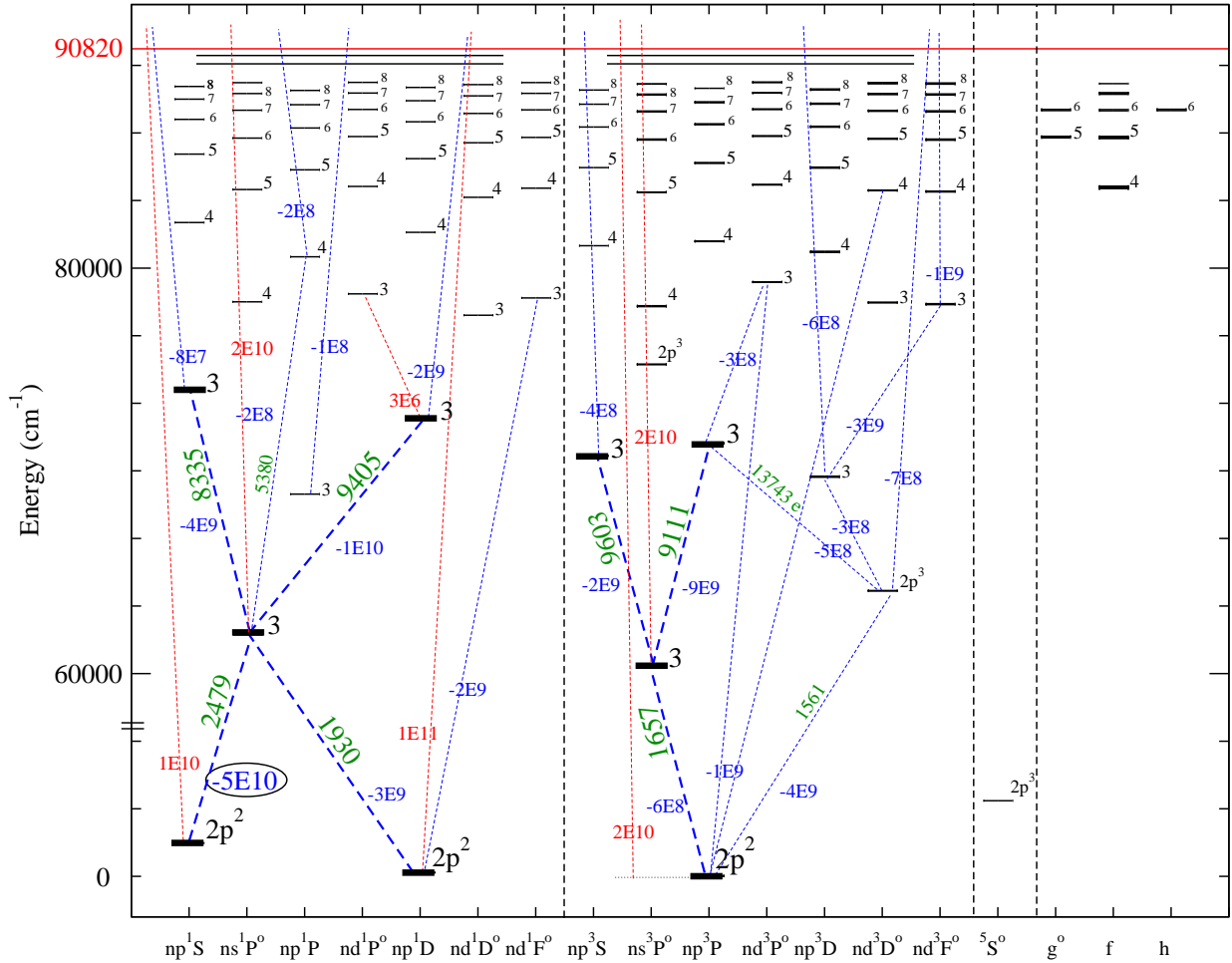}
\caption{The NET-diagram at log$\tau_{5000} = -1.3$ in the model 10400/3.5/0. Wavelengths (\,\AA\ ) and NET values  (\scm) are quoted for the selected transitions. NETs of negative sign are marked in blue, and positive marked in red (only on-line available).} 
\label{NET}
%\end{center}
\end{figure*}

Which processes in C\ione\ do drive the emission in the IR lines? How do they depend on atmospheric parameters? 
In Fig.\,\ref{NET}, we present the NET rates, NET = $n_l (R_{lu} + C_{lu}) - n_u (R_{ul} + C_{ul})$, in the model 10400/3.5/0 at log$\tau_{5000} = -1.3$, where the C\ione\ 8335 and 9405\,\AA\ emission form (Fig.\,\ref{DC}). Here, $R_{lu}$ and $C_{lu}$ are radiative and collisional rates, respectively, for the transition from low to upper level, and $R_{ul}$ and $C_{ul}$ for the inverse transition. The high-excitation levels, with \Eexc\ $> $ 7.9~eV, including the upper levels of the investigated transitions, are mainly populated by recombination from C\ii\ that follows an overionization of the C\ione\ ground state (NET = $2\cdot 10^{10}$  \scm) and low-excitation levels, including the lower levels of the investigated transitions (NET varies between $1\cdot 10^{10}$ \scm and $1\cdot 10^{11}$ \scm). Thus, the overionization-recombination mechanism resulting in a depopulation of the lower levels to a greater extent than the upper levels is a prerequisite of the emission phenomenon.
 Extra depopulation of the levels 3s$^{1}$P$^{\circ}$ and 3s$^{3}$P$^{\circ}$ can be caused by photon loss in the UV transitions 2p$^{2}$~$^{1}$S -- 3s$^{1}$P$^{\circ}$ (2479\,\AA), 2p$^{2}$~$^{1}$D -- 3s$^{1}$P$^{\circ}$ (1930\,\AA), and 2p$^{2}$~$^{3}$P -- 3s$^{3}$P$^{\circ}$ (1657\,\AA). An influence of these UV transitions on the SE of C\ione\ in the layers, where the IR lines form, depends on atmospheric parameters. In the model 10400/3.5/0, photon loss in C\ione\ 2479\,\AA\ drains population of 3s$^{1}$P$^{\circ}$ effectively in the layers around log$\tau_{5000} = -1.3$, with NET = $-5\cdot 10^{10}$ \scm, while the transitions corresponding to C\ione\ 1930\,\AA\ and 1657\,\AA\ are in detailed balance because of $\tau_{1930} \gg 1$ and $\tau_{1657} \gg 1$. 
We performed test calculations by reducing a radiative rate of C\ione\ 2479\,\AA\ by 10\,\%. This resulted in weakened emission, such that the line centre relative flux of C\ione\ 9405\,\AA\ reduced from 1.053 to 1.028.  

With increasing \Teff, optical depth of formation of the C\ione\ UV lines shift inwards (Fig.\,\ref{DC} for \Teff\ = 17\,500~K) and photon losses depopulate not only 3s$^{1}$P$^{\circ}$, but also 3s$^{3}$P$^{\circ}$, resulting in the emission triplet lines at 9061 -- 9111\,\AA\ and 9603 -- 9658\,\AA.

\section{Carbon lines in the selected stars}\label{Sect:Stars}

\begin{table*}
   \begin{normalsize}
  \caption{ Lines of C\ione\ and C\ii\ used in abundance analysis. }
          \label{tab1} 
  \begin{tabular}{lccccccc}\hline  \hline                                                   
$\lambda$, \AA\ & Transition   &                          log~$gf$ & \Eexc, eV  & $\lambda$, \AA\ & Transition   &                          log~$gf$ & \Eexc, eV   \\\hline
\multicolumn{4}{c}{C\ione\ } & \multicolumn{4}{c}{C\ione\ }\\                                                         
1329.09 & 2p$^{2}$ $^{3}$P$_{1}$ -- 2p$^{3}$ $^{3}$P$_{0}^{\circ}$ &  $-$1.231 &  0.00    & 9094.83 & 3s $^{3}$P$_{2}^{\circ}$ -- 3p $^{3}$P$_{2}$           & 0.151       &  7.49    \\
1329.10 & 2p$^{2}$ $^{3}$P$_{1}$ -- 2p$^{3}$ $^{3}$P$_{2}^{\circ}$ &  $-$1.147 &  0.00    & 9111.80 & 3s $^{3}$P$_{2}^{\circ}$ -- 3p $^{3}$P$_{1}$           & $-$0.297    &  7.49     \\ 
1329.12 & 2p$^{2}$ $^{3}$P$_{1}$ -- 2p$^{3}$ $^{3}$P$_{1}^{\circ}$ &  $-$1.355 &  0.00    & 9603.02  & 3s $^{3}$P$_{0}^{\circ}$ -- 3p $^{3}$S$_{1}$           & $-$0.896    &  7.48    \\ 
1329.59 & 2p$^{2}$ $^{3}$P$_{1}$ -- 2p$^{3}$ $^{3}$P$_{2}^{\circ}$ &  $-$0.662 &  0.01    & 9658.43 & 3s $^{3}$P$_{2}^{\circ}$ -- 3p $^{3}$S$_{1}$           & $-$0.280    &  7.49     \\
1329.60 & 2p$^{2}$ $^{3}$P$_{1}$ -- 2p$^{3}$ $^{3}$P$_{2}^{\circ}$ &  $-$1.136 &  0.01    & 10123.87 & 3p $^{1}$P$_{1}$ -- 3d $^{1}$P$_{1}^{\circ}$           & $-$0.03     &  8.54    \\
1459.03 & 2p$^{2}$ $^{1}$D$_{2}$ -- 3d $^{1}$P$_{1}^{\circ}$       &  $-$1.282 &  1.26    & 10691.24 & 3s $^{3}$P$_{2}^{\circ}$ -- 3p $^{3}$D$_{3}$           &  0.344      &  7.49     \\
1463.34 & 2p$^{2}$ $^{1}$D$_{2}$ -- 3d $^{1}$F$_{3}^{\circ}$       &  $-$0.396 &  1.26    & 10683.08 & 3s $^{3}$P$_{1}^{\circ}$ -- 3p $^{3}$D$_{2}$           &  0.08       &  7.48      \\ 
1657.91 & 2p$^{2}$ $^{3}$P$_{1}$ -- 3s $^{3}$P$_{0}^{\circ}$       &  $-$0.845 &  0.00    & 10685.36 & 3s $^{3}$P$_{0}^{\circ}$ -- 3p $^{3}$D$_{1}$           & $-$0.27     &  7.48      \\ 
1658.12 & 2p$^{2}$ $^{3}$P$_{2}$ -- 3s $^{3}$P$_{1}^{\circ}$       &  $-$0.748 &  0.01    & 10729.53 & 3s $^{3}$P$_{2}^{\circ}$ -- 3p $^{3}$D$_{2}$           & $-$0.42     &  7.49      \\
4762.52 & 3s $^{3}$P$_{1}^{\circ}$ -- 4p $^{3}$P$_{2}$           &  $-$2.335   &  7.48    & 10707.32 & 3s $^{3}$P$_{1}^{\circ}$ -- 3p $^{3}$D$_{1}$           & $-$0.41     &  7.48      \\
4766.66 & 3s $^{3}$P$_{1}^{\circ}$ -- 4p $^{3}$P$_{1}$           &  $-$2.617   &  7.48    & 10753.98 & 3s $^{3}$P$_{2}^{\circ}$ -- 3p $^{3}$D$_{1}$           & $-$1.60     &  7.49   \\
4770.02 & 3s $^{3}$P$_{1}^{\circ}$ -- 4p $^{3}$P$_{0}$           &  $-$2.437   &  7.48    &   \multicolumn{4}{c}{C\ii\ }       \\  
4771.73 & 3s $^{3}$P$_{2}^{\circ}$ -- 4p $^{3}$P$_{2}$           &  $-$1.866   &  7.49    &  1335.70  &  2p $^{2}$P$_{3/2}^{\circ}$ -- 2p$^{2}$ $^{2}$D$_{5/2}$       & $-$0.335 & 0.01   \\
4775.89 & 3s $^{3}$P$_{2}^{\circ}$ -- 4p $^{3}$P$_{1}$           &  $-$2.304   &  7.49    &  1323.86  &  2p$^{2}$ $^{2}$D$_{5/2}$  --  2p$^{3}$ $^{2}$D$_{3/2}^{\circ}$  & $-$1.284 & 9.29  \\  
4932.04 & 3s $^{1}$P$_{1}^{\circ}$ -- 4p$^{1}$S$_{0}$            &  $-$1.658   &  7.69    &  1323.91  &  2p$^{2}$ $^{2}$D$_{3/2}$  --  2p$^{3}$ $^{2}$D$_{3/2}^{\circ}$  & $-$0.337 & 9.29  \\ 
5052.14 & 3s $^{1}$P$_{1}^{\circ}$ -- 4p $^{1}$D$_{2}$           &  $-$1.303   &  7.69    &  1323.95  &  2p$^{2}$ $^{2}$D$_{5/2}$  --  2p$^{3}$ $^{2}$D$_{5/2}^{\circ}$  & $-$0.144 & 9.29  \\
5380.32 & 3s $^{1}$P$_{1}^{\circ}$ -- 4p $^{1}$P$_{1}$           &  $-$1.616   &  7.69    &  1324.00  &  2p$^{2}$ $^{2}$D$_{3/2}$  --  2p$^{3}$ $^{2}$D$_{5/2}^{\circ}$  & $-$1.288 & 9.29  \\
6007.17 & 3p $^{3}$D$_{1}$ -- 6s $^{3}$P$_{1}^{\circ}$           &  $-$2.062   &  8.64    &  3918.96  &  3p $^{2}$P$_{1/2}^{\circ}$ -- 4s $^{2}$S$_{1/2}$       & $-$0.533  &   16.33  \\
6012.22 & 3p $^{3}$D$_{1}$ -- 5d $^{3}$F$_{2}^{\circ}$           &  $-$2.005   &  8.64    &  3920.68  &  3p $^{2}$P$_{3/2}^{\circ}$ -- 4s $^{2}$S$_{1/2}$       & $-$0.232  &   16.33  \\
6013.21 & 3p $^{3}$D$_{3}$ -- 5d $^{3}$F$_{4}^{\circ}$           &  $-$1.314   &  8.65    &  4267.00  &  3d $^{2}$D$_{3/2}$ -- 4f $^{2}$F$_{5/2}^{\circ}$       & 0.563     &   18.05  \\
6014.83 & 3p $^{3}$D$_{2}$ -- 6s $^{3}$P$_{1}^{\circ}$           &  $-$1.584   &  8.64    &  4267.26  &  3d $^{2}$D$_{5/2}$ -- 4f $^{2}$F$_{7/2}^{\circ}$       & 0.716     &   18.05  \\
6587.61 & 3p $^{1}$P$_{1}$ -- 4d $^{1}$P$_{1}^{\circ}$           &  $-$1.003   &  8.54    &  4267.26  &  3d $^{2}$D$_{5/2}$ -- 4f $^{2}$F$_{5/2}^{\circ}$       & $-$0.584  &   18.05  \\ 
7111.46  & 3p $^{3}$D$_{1}$ -- 4d $^{3}$F$_{2}^{\circ}$           &  $-$1.09   &  8.64    &  5132.95  & 2p3s 4P$_{1/2}^{\circ}$ -- 2p3p 4P$_{3/2}$  &  $-$0.211 & 20.70\\ 
7113.17  & 3p $^{3}$D$_{3}$ -- 4d $^{3}$F$_{4}^{\circ}$           &  $-$0.77   &  8.65    &  5133.28  & 2p3s 4P$_{3/2}^{\circ}$ -- 2p3p 4P$_{5/2}$  &  $-$0.178 & 20.70\\  
7115.17  & 3p $^{3}$D$_{2}$ -- 4d $^{3}$F$_{3}^{\circ}$           &  $-$0.93   &  8.64    &  5137.25  & 2p3s 4P$_{1/2}^{\circ}$ -- 2p3p 4P$_{1/2}$  &  $-$0.911 & 20.70\\  
7115.18  & 3p $^{3}$D$_{1}$ -- 5s $^{3}$P$_{0}^{\circ}$           &  $-$1.47   &  8.64    &  5139.17  & 2p3s 4P$_{3/2}^{\circ}$ -- 2p3p 4P$_{3/2}$  &  $-$0.707 & 20.70\\  
7116.98  & 3p $^{3}$D$_{3}$ -- 5s $^{3}$P$_{2}^{\circ}$           &  $-$0.91   &  8.65    &  5143.49  & 2p3s 4P$_{3/2}^{\circ}$ -- 2p3p 4P$_{1/2}$  &   $-$0.212 &   20.70\\  
7119.65  & 3p $^{3}$D$_{2}$ -- 5s $^{3}$P$_{1}^{\circ}$           &  $-$1.148  &  8.64    &  5145.16  & 2p3s 4P$_{5/2}^{\circ}$ -- 2p3p 4P$_{5/2}$  &   0.189    &   20.71\\  
8335.14  & 3s $^{1}$P$_{1}^{\circ}$ -- 3p $^{1}$S$_{0}$           &  $-$0.437  &  7.69    &  5151.09  & 2p3s 4P$_{5/2}^{\circ}$ -- 2p3p 4P$_{3/2}$  &   $-$0.179 &   20.71\\  
9405.73 & 3s $^{1}$P$_{1}^{\circ}$ -- 3p $^{1}$D$_{2}$           &  0.286      &  7.69    &  6578.05  & 3s $^{2}$S$_{1/2}$ -- 3p $^{2}$P$_{3/2}^{\circ}$       & $-$0.021  &   14.45  \\  
9061.43 & 3s $^{3}$P$_{1}^{\circ}$ -- 3p $^{3}$P$_{2}$           & $-$0.347    &  7.48    &  6582.88  & 3s $^{2}$S$_{1/2}$ -- 3p $^{2}$P$_{1/2}^{\circ}$       & $-$0.323  &   14.45  \\   
9062.49 & 3s $^{3}$P$_{0}^{\circ}$ -- 3p $^{3}$P$_{1}$           & $-$0.455    &  7.48    &  7231.33  & 3p $^{2}$P$_{1/2}^{\circ}$ - 3d $^{2}$D$_{3/2}$        &   0.039   &   16.33  \\   
9078.28 & 3s $^{3}$P$_{1}^{\circ}$ -- 3p $^{3}$P$_{1}$           & $-$0.581    &  7.48    &  7236.41  & 3p $^{2}$P$_{3/2}^{\circ}$ - 3d $^{2}$D$_{5/2}$        &  0.294    &   16.33  \\   
9088.51  & 3s $^{3}$P$_{1}^{\circ}$ -- 3p $^{3}$P$_{0}$           & $-$0.43    &  7.48    &  7237.16  & 3p $^{2}$P$_{3/2}^{\circ}$ - 3d $^{2}$D$_{3/2}$        & $-$0.660  &   16.33 \\  \hline                                 %\multicolumn{5}{l}{Mult.: the multiplet numbers accordingly to \citet{1972mtai.book.....M}.} \\ 
  \end{tabular}
\end{normalsize}
\end{table*}  

\subsection{Stellar sample, observations, and atmospheric parameters }   

Our sample includes seven sharp-lined stars, with well-determined stellar parameters (Table~\ref{tab3}). Due to low rotational velocities, with $V$sin$i$ $\lesssim$~40~\kms, they allow spectral analysis to be done at the highest precision, maximizing the chance to identify even weak emission lines (WELs) in their spectra.
They are either single stars, or primaries in close binary systems (SB1) with much fainter
companions, or individual components in a visual binary. 

\begin{table*}
 \begin{minipage}{150mm}
 \begin{normalsize}
     \caption{Atmospheric parameters of the selected stars and sources of the data.}
        \label{tab3} 
  \begin{tabular}{rcccccccc}\hline 
  \hline 
HD      &    Name      &   Sp. t.    &   \Teff  &  log~$g$&    [Fe/H] &    $\xi_t$ &   $V$sin$i$  &     Ref.    \\         
        &              &             &    (K)   &  (CGS) &           &  (\kms)    &  (\kms)    &                \\\hline
17081   &  $\pi$ Cet   &   B7 IV E   &   12800  &  3.8   &   0.0     &   0.5      & 20         &  1            \\
22136   &     --       &   B8V C     &   12700  &  4.2   &  $-$0.28  &   1.1      & 15         &  2             \\
48915   &   Sirius     &   A1V+DA    &   9850   &  4.3   &   0.4     &   1.8      & 16.5       &  3             \\
73666   &   40 Cnc     &   A1V C     &   9382   &  3.78  &   0.16    &   1.9      & 10         &  4          \\
160762  &  $\iota$ Her &   B3 IV SPB &   17500  &  3.8   &   0.02    &   1.0      & 6          &  5             \\
172167  &   Vega       &   A0Va C    &   9550   &  3.95  &   $-$0.5  &   2.0      & 14         &  6            \\ 
209459  &   21 Peg     &   B9.5V C   &   10400  &  3.5   &   0.0     &   0.5      & 4          &  1            \\ \hline
  \end{tabular}                                                                                                                                                   
\begin{tablenotes}                                                                                                                                                     
\item[a] {\bf References:} 1 = \citet{2009A&A...503..945F};
2 = \citet{2013A&A...551A..30B}; 3 = \citet{1993AA...276..142H}; 4 = \citet{2007A&A...476..911F}; 5 = \citet{2012A&A...539A.143N}; 6 = \citet{2000A&A...359.1085P}.
\end{tablenotes}  
%\end{center}
 \end{normalsize}
   \end{minipage}
\end{table*}   
   
\begin{table*}
 \begin{minipage}{130mm}
%     \caption{Characteristics of observed spectra. }
     \caption{Characteristics of observed spectra. }
        \label{tab2} 
  \begin{tabular}{@{}lcccccccc}\hline 
  \hline 
HD          & V$^1$    & Telescope/        &  Spectral range& $t_{exp}$     & Observing run& $R$     &   S/N  \\
            & (mag)    & spectrograph      &  (\AA\,)          & (s)    & year/month  &         &       \\ \hline       
17081       &  4.2     &  1                 & 3690$-$10480    & 120     & 2005/02     & 65000   & 600 \\
22136       &  6.9     &  1                 & 3690$-$10480    & 356     & 2008/03     & 65000   & 500 \\
48915       &  $-$1.5  &  1                 & 3690$-$10480    & 0.6     & 2011/02     & 65000   & 500 \\
            &          &  2                 & 1265$-$1368     & 3482     & 1996/11     & 25000   & 200 \\
73666       &  6.6     &  1                 & 3690$-$10480    & 1600     & 2006/01     & 65000   & 660 \\
160762      &  3.8     &  1                 & 3690$-$10480    & 240     & 2012/06     & 65000   & 600 \\
172167      &  0.0     &  1                 & 3690$-$10480    & 8     & 2011/07     & 65000   & 500 \\
209459      &  5.8     &  1                 & 3690$-$10480    & 290    & 2013/08     & 65000   & 600 \\ \hline
  \end{tabular}
  \begin{tablenotes}  
\item[a] {\bf Notes.} $^1$ V is a visual magnitude from the SIMBAD database. \\ {\bf Telescope / spectrograph:} 1 = CFHT/ESPaDOnS; 2 = HST/GHRS. 
\end{tablenotes}  
 \end{minipage}
\end{table*}

For all the stars, the spectra in the visible spectral range (3690--10480\,\AA) were obtained with the Echelle SpectroPolarimetric Device
for the Observation of Stars (ESPaDOnS) attached at the 3.6 m telescope of the
Canada--France--Hawaii Telescope (CFHT) observatory. Spectra were extracted from the ESPaDOnS archive\footnote{http://www.cfht.hawaii.edu/Instruments/Spectroscopy/Espadons/}.
The resolving power and signal-to-noise ratio are R = 65\,000 and S/N $\simeq$ 500-600, respectively. 
For Sirius, we also used the UV spectrum that covers the wavelength region between 1265 and 1368\,\AA. 
It was obtained with the Goddard High Resolution Spectrograph on the $Hubble Space Telescope$ ($HST$), and it has R = 25\,000 and S/N of 100--200, in general.
Characteristics of the observed spectra for individual stars are given in Table~\ref{tab2}.

We comment on the programme stars. 

The star 21~Peg (HD~209459) is known from previous studies as a 'normal' single star with
$V$sin$i$ $\sim$ 4~\kms\ \citep{1981PASP...93..587S, 2009A&A...503..945F}.
 
The star $\pi$~Cet (HD~17081) is an SB1 star with $V$sin$i$~$\sim$~20~\kms. The observed small infrared excess at $\lambda \leq 3\mu$ makes it a candidate for Herbig Ae/Be star \citep{1998A&A...331..211M}. Since its spectrum is not
visibly contaminated by the companion, it fully serves for our analysis. This star was already used  by \citet{1993A&A...274..335S} as a normal comparison star in the abundance study of chemically peculiar stars.

For 21~Peg and $\pi$~Cet, their effective temperatures and surface
gravities were spectroscopically determined from the Balmer lines \citep{2009A&A...503..945F}.

HD~22136 is also a normal star with $V$sin$i$ $\sim$ 15~\kms, showing no chemical peculiarities. Its fundamental parameters were derived by \citet{2013A&A...551A..30B} based on the Geneva and uvby$\beta$ photometry.

Effective temperature and surface
gravity of Vega (HD~172167) were determined by \citet{2000A&A...359.1085P} from the Balmer line wings and Mg\ione\ and Mg\ii\ lines.
Vega is a rapidly rotating star seen pole-on. Rapid rotation causes a change of the surface shape from spherical to ellipsoidal one, therefore, the temperature stratification and surface gravity vary from the pole to equator \citep{2010ApJ...712..250H}. We ignore the non-spherical effects in our study and 
analyse Vega's flux spectrum using the average temperature and gravity.  
Vega was also classified as a mild $\lambda$~Bootis-type star \citep{1990ApJ...363..234V}.

$\iota$~Her (HD~160762) is a single bright star, with $V$sin$i$ $\sim$ 6~\kms. Its atmospheric parameters were
determined spectroscopically from all available H and He lines and multiple
ionisation equilibria, and they were confirmed via the spectral energy distribution and the $Hipparcos$ distance \citep{2012A&A...539A.143N}.

HD~73666 is a Blue Straggler and a member of the Praesepe cluster.
It is previously considered as an Ap (Si) star, but appears to have the abundances of a normal A-type star \citep{2007A&A...476..911F}.
This star is a primary component of SB1, as is the case for many other Blue Stragglers \citep{1996ASPC...90..337L}.
The flux coming from the secondary star is negligible, as previously checked by \citet{1998A&A...338.1073B}, so the star
was analysed ignoring the presence of the secondary. The physical parameters were taken from \citet{2007A&A...476..911F}. 
Both Fe\ione\ excitation (for the effective temperature) and Fe\ione/Fe\ii\ ionisation (for log~$g$) equilibria were used in the parameter determination. 

The Sirius binary system (HD 48915 = HIP 32349) is composed of a main sequence A1V star and a hot DA white dwarf.
It is an astrometric visual binary system at a distance of only 2.64~pc. Their physical parameters are available with the 
highest precision and were taken from \citet{1993AA...276..142H}. Sirius~A is classified as a hot metallic-line (Am) star.

\begin{figure*}
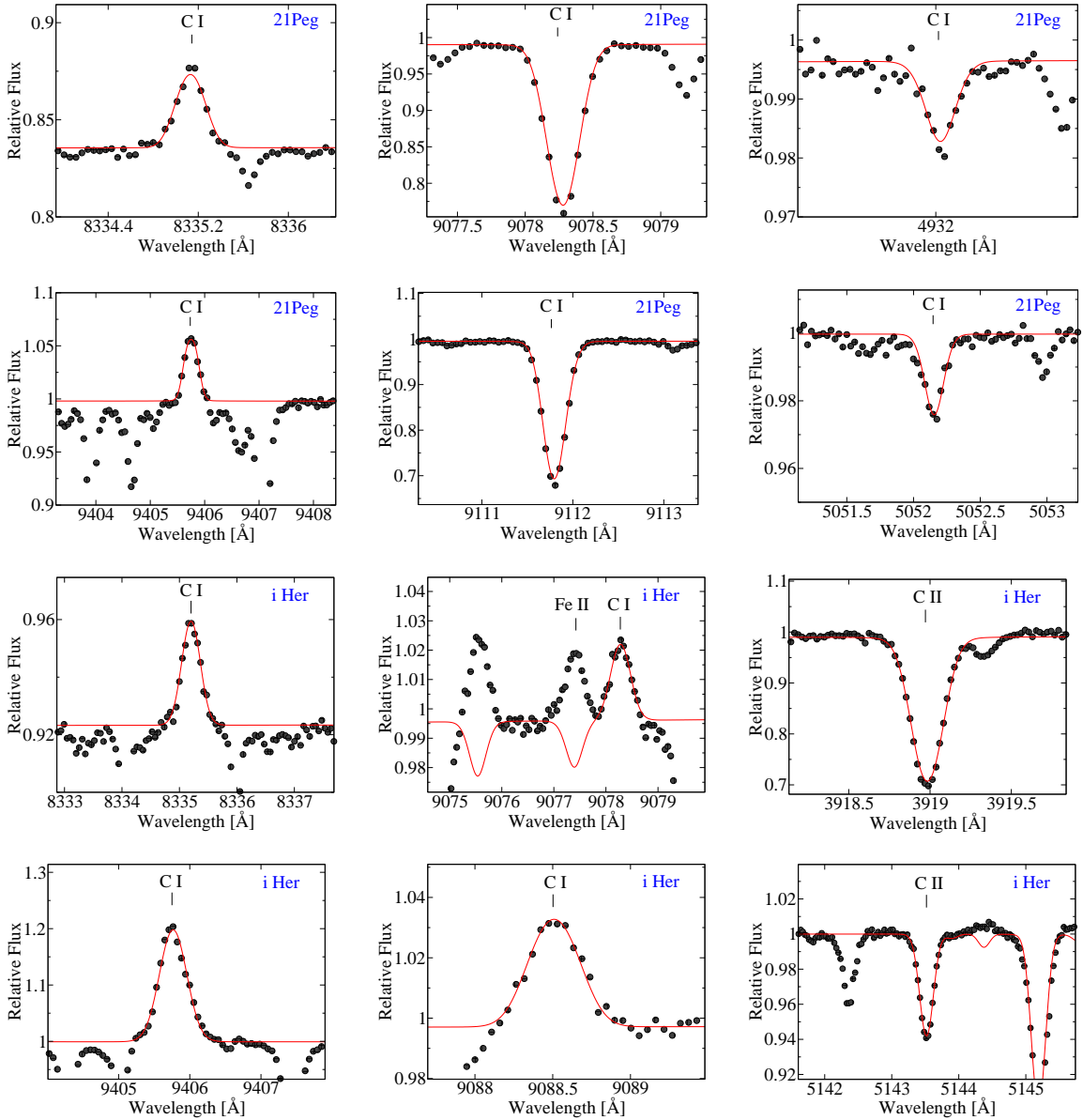

\begin{minipage}{150mm}
%\begin{center}
\parbox{0.3\linewidth}{\includegraphics[scale=0.17]{Figure18.eps}\\
\centering}
%\hspace{0.33\linewidth}
\parbox{0.3\linewidth}{\includegraphics[scale=0.17]{Figure19.eps}\\
\centering}
\parbox{0.3\linewidth}{\includegraphics[scale=0.17]{Figure20.eps}\\
\centering}
\hspace{1\linewidth}
\hfill
\\[0ex]
%\hspace{0.33\linewidth}
\parbox{0.3\linewidth}{\includegraphics[scale=0.17]{Figure21.eps}\\
\centering}
\parbox{0.3\linewidth}{\includegraphics[scale=0.17]{Figure22.eps}\\
\centering}
\parbox{0.3\linewidth}{\includegraphics[scale=0.17]{Figure23.eps}\\
\centering}
\hspace{1\linewidth}
\hfill
\\[0ex]
%\hspace{0.33\linewidth}
\parbox{0.3\linewidth}{\includegraphics[scale=0.17]{Figure24.eps}\\
\centering}
\parbox{0.3\linewidth}{\includegraphics[scale=0.17]{Figure25.eps}\\
\centering}
\parbox{0.3\linewidth}{\includegraphics[scale=0.17]{Figure26.eps}\\
\centering}
\hspace{1\linewidth}
\hfill
\\[0ex]
%\hspace{0.33\linewidth}
\parbox{0.3\linewidth}{\includegraphics[scale=0.17]{Figure27.eps}\\
\centering}
\parbox{0.3\linewidth}{\includegraphics[scale=0.17]{Figure28.eps}\\
\centering}
\parbox{0.3\linewidth}{\includegraphics[scale=0.17]{Figure29.eps}\\
\centering}
\hspace{1\linewidth}
\hfill
\\[0ex]
\caption{Best fits (solid curves) of the selected C\ione\ and C\ii\ lines in 
21~Peg (two top rows) and $\iota$~Her (two bottom rows).
The observed spectra are shown by bold dots. 
 }
\label{pics}
\end{minipage}
\end{figure*} 

\subsection{Analysis of C\ione\ emission lines in B-type stars}\label{Sect:emission}

Emission lines of C\ione\ were detected in the observed spectra of the four stars: 21~Peg, HD~22136, $\pi$~Cet, and $\iota$~Her. 
In 21~Peg, HD~22136, and $\pi$~Cet only C\ione\ 8335 and 9405\,\AA\ are in emission,   while the hottest star of our sample, $\iota$~Her, also has the emission lines at 9061--9111\,\AA\ and 9603--9658\,\AA. Some examples are shown in Fig.\,\ref{pics}.

\begin{figure}
\includegraphics[scale=0.3]{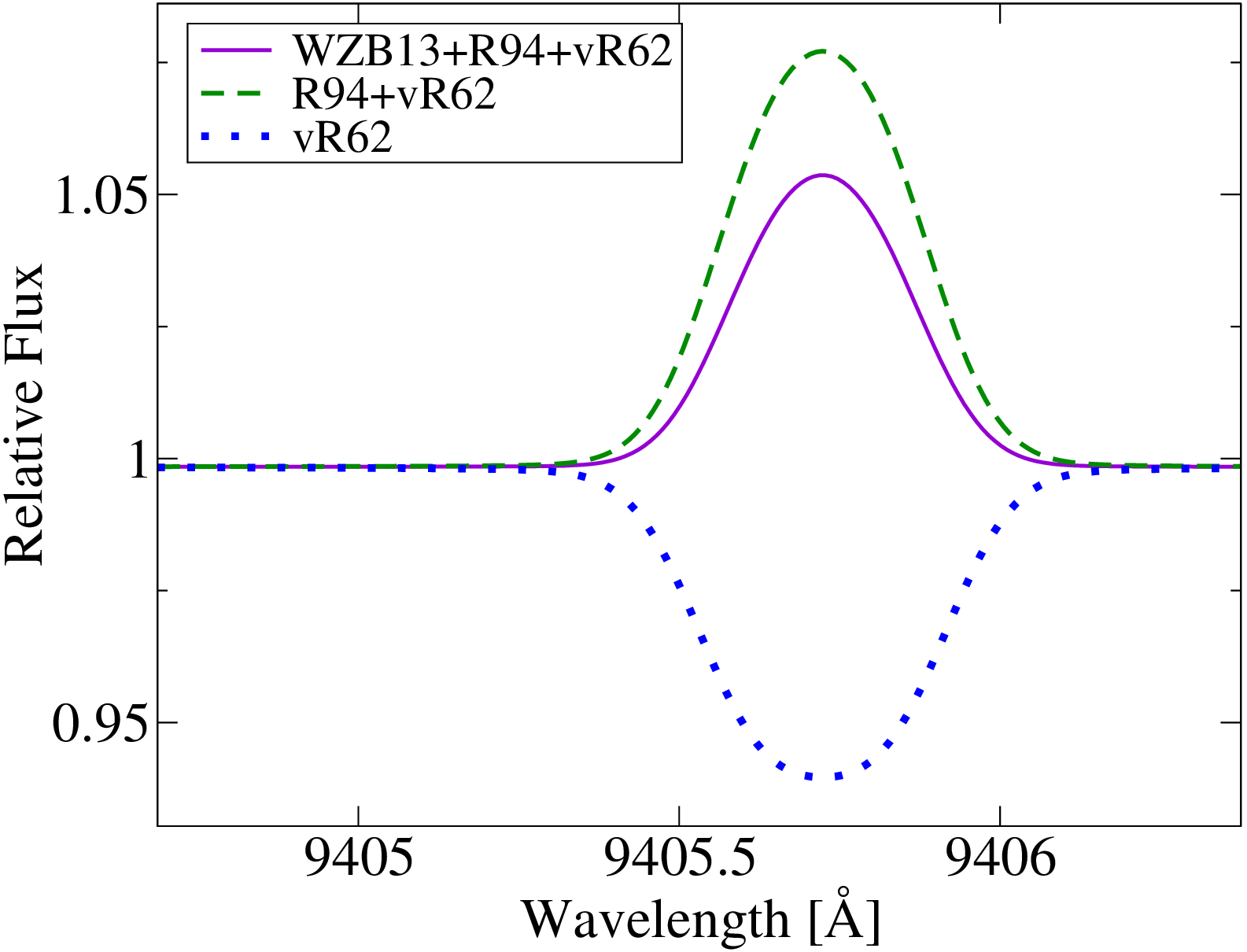}
\caption{NLTE line profiles of C\ione\ 9405\,\AA\ in the model 10400/3.5/0 from calculations with collisional data of \citet{2013PhRvA..87a2704W} where available, \citet{Reid1994}, and  \citet{1962ApJ...136..906V} (solid curve, WZB13+R94+vR62), 
collisional data from \citet{Reid1994} for 537 transitions and formula of \citet{1962ApJ...136..906V} for the remaining ones (dashed curve, R94+vR62), and using formula of \citet{1962ApJ...136..906V} only (dotted curve, vR62). Everywhere, log$\epsilon_{\rm C}$ = 8.43. The theoretical spectra are convolved with an instrumental profile of R = 65\,000.}
\label{Zat}
%\end{center}
\end{figure}

For each star, our NLTE calculations reproduce well the observed C\ione\ emission lines, using the treated model atom of C\ione\ -- C\ii\ and adopted atmospheric parameters. The best NLTE fits of the selected emission lines in 21~Peg and $\iota$~Her are presented in Fig.\,\ref{pics}.

Analysis of stellar emission lines provides an opportunity to test collisional data for C\ione. 
As described in Sect.\,\ref{Sect:modelatom}, electron-impact excitation data for C\ione\ were computed by \citet{2013PhRvA..87a2704W} and \citet{Reid1994}. We checked the influence of varying collisional rates on the appearance of the C\ione\ emission lines in the model 10400/3.5/0 representing the atmosphere of 21~Peg. 
Calculations were performed with the three different sets of data, 
namely ($i$) 703 transitions from \citet{2013PhRvA..87a2704W} plus 323 transitions from \citet{Reid1994} 
and plus formula of \citet{1962ApJ...136..906V} for the remaining transitions; 
this is our standard model as described in Sect.\,\ref{Sect:modelatom}, ($ii$)  537 transitions from \citet{Reid1994} 
 and the formula of \citet{1962ApJ...136..906V} and $\Omega$ = 1 
for the remaining allowed and forbidden transitions, 
($iii$) only the formula of \citet{1962ApJ...136..906V} and $\Omega$ = 1 were employed. 
Fig.\,\ref{Zat} shows that the effect on C\ione\ 9405\,\AA\ is large. With the theoretical approximations, we computed the absorption line profile, and this is not supported by observations of 21~Peg, where C\ione\ 9405\,\AA\ is in emission. Only applying accurate collisional rates allows us to obtain the emission. However, the emission is stronger in calculations with \citet{Reid1994} than \citet{2013PhRvA..87a2704W} data. The choice was decided by comparison with C\ione\ 9405\,\AA\ observed in 21~Peg. With the element abundance determined from the C\ione\ absorption lines, a strength of the C\ione\ 9405\,\AA\ emission was reproduced, when using data of \citet{2013PhRvA..87a2704W}. They were employed everywhere in further stellar abundance analysis.

\begin{table*}
% \begin{center}
\begin{minipage}{130mm}
  \caption{NLTE abundances from C\ione\ lines in Vega and NLTE abundance corrections.  }
          \label{tab4} 
  \begin{tabular}{lcccccccccc}\hline  \hline                                                   
 &  \multicolumn{3}{c}{ WZB13+R94+vR62}&   \multicolumn{1}{c}{ R94+vR62}  & \multicolumn{1}{c}{ vR62} &P2001 & SH1990\\ 
$\lambda$, \AA & NLTE & LTE & $\Delta_{\rm NLTE}$ & NLTE / $\Delta_{\rm NLTE}$ & NLTE / $\Delta_{\rm NLTE}$ & NLTE & NLTE  \\\hline 
4771.73   &  8.52 & 8.56 & -0.04  & 8.53 / -0.03       &   8.53 / -0.03   &   8.38   &   8.53\\
4775.89   &  8.48 & 8.52 & -0.04  & 8.49 / -0.03       &   8.50 / -0.02   &   8.35   &       \\
5052.14   &  8.23 & 8.27 & -0.04  & 8.24 / -0.03       &   8.25 / -0.02   &   8.14   &   8.23\\      
5380.32   &  8.28 & 8.34 & -0.06  & 8.29 / -0.05       &   8.30 / -0.04   &   8.19   &   8.30\\   
6587.61   &  8.26 & 8.33 & -0.07  & 8.27 / -0.06       &   8.28 / -0.05   &   8.20   &       \\  
7111.46   &  8.26 & 8.32 & -0.06  & 8.27 / -0.05       &   8.28 / -0.04   &   8.14   &   8.24\\    
7113.17   &  8.12 & 8.18 & -0.06  & 8.13 / -0.05       &   8.14 / -0.04   &   8.03   &   8.10\\   
7116.98   &  8.28 & 8.34 & -0.06  & 8.29 / -0.05       &   8.30 / -0.04   &   8.17   &   8.27\\   
7115.17\multirow{2}{*}{$\Big\}$} & \multirow{2}{*}{8.20} & \multirow{2}{*}{8.26} & \multirow{2}{*}{-0.06} &  \multirow{2}{*}{8.21 / -0.05}     &   \multirow{2}{*}{8.22 / -0.04}  &  \multirow{2}{*}{8.10 } &  \multirow{2}{*}{8.28} \\   
7115.18   &                   &                     &                 &                \\\hline 
Mean(vis) &  8.32 & 8.36 &     &  8.33   &   8.34 &   8.21 & 8.30  \\ 
$\sigma$  &  0.13 & 0.12 &     &  0.13   &   0.13 &   0.12 & 0.14  \\ \hline 
9078.28   &  8.39 & 8.84 & -0.45  & 8.39 / -0.45    &  8.56 / -0.28   &  8.23  &      \\   
9088.51   &  8.35 & 8.88 & -0.53  & 8.36 / -0.52    &  8.61 / -0.27   &  8.40  &  8.33\\   
9111.80   &  8.27 & 8.86 & -0.59  & 8.29 / -0.57    &  8.49 / -0.37   &  8.31  &  8.35\\ \hline 
Mean(IR)  & 8.34  & 8.86 &        &  8.35           &   8.56          &  8.32  & 8.34 \\ 
$\sigma$  & 0.06  & 0.02 &        &  0.05           &   0.06          &  0.09  & 0.01 \\\hline 
Mean(vis+IR)& 8.33 & 8.55&        & 8.34           &   8.41          & 8.25 & 8.31 \\   
$\sigma$  & 0.11 & 0.25  &        &  0.11          &   0.16          & 0.12 & 0.11 \\\hline  
10123.87  & 8.10 & 8.38 &-0.28    &  8.12 / -0.26  &   8.26 / -0.12  &   &  8.21 \\  
10683.08  & 8.45 & 8.99 &-0.54    &  8.43 / -0.56  &   8.71 / -0.28  &   &  8.45 \\   
10685.36  & 8.62 & 9.03 &-0.41    &  8.67 / -0.36  &   8.88 / -0.15  &   &  8.56 \\   
10691.24  & 8.26 & 8.90 &-0.64    &  8.24 / -0.66  &   8.53 / -0.37  &   &  8.36 \\   
10729.53  & 8.16 & 8.52 &-0.36    &  8.12 / -0.40  &   8.26 / -0.26  &   &  8.22 \\   
10707.32  & 8.32 & 8.68 &-0.36    &  8.28 / -0.40  &   8.42 / -0.26  &   &  8.33 \\
10753.98  & 8.35 & 8.44 &-0.09    &  8.37 / -0.07  &   8.45 /  0.01  &   &  8.33 \\ \hline 
Mean(all) & 8.34 & 8.65 &         &  8.35          &   8.47          &8.25 & 8.34 \\ 
$\sigma$  & 0.13 & 0.28 &         &  0.13          &   0.20          &0.12 & 0.12 \\\hline 
  \end{tabular}
  \begin{tablenotes}  
\item {\bf Notes:} P2001: \citet{2001AA...379..936P}, SH1990: \citet{1990AA...237..125S}.
\end{tablenotes}  
 \end{minipage}
\end{table*} 

\begin{table}
% \begin{minipage}{130mm}
     \caption{ Carbon NLTE abundances of Sirius}
        \label{tab5} 
  \begin{tabular}{@{}lcccccccc}\hline 
  \hline 
$\lambda$, \AA\ &  EW, m\AA\  &  NLTE & LTE & $\Delta_{\rm NLTE}$ \\\hline
C\ione\      &           &    &    &           \\
1329.09\multirow{3}{*}{$\bigg\}$}  &  \multirow{3}{*}{bl}      & \multirow{3}{*}{7.90} & \multirow{3}{*}{7.89} &  \multirow{3}{*}{+0.01}    \\  
1329.10  &         &    &      &         \\
1329.12  &         &    &      &         \\                                    
1329.59\multirow{2}{*}{$\Big\}$}  &  \multirow{2}{*}{bl}     & \multirow{2}{*}{7.79} & \multirow{2}{*}{7.79} &  \multirow{2}{*}{0.00}    \\ 
1329.60  &        &         &      &            \\                  
1459.03  &  bl    & 7.84    & 7.84 &    0.00     \\  
1463.34  &  bl    & 7.67    & 7.66 &   +0.01     \\ 
1657.91  &  bl    & 7.45    & 7.44 &   +0.01     \\ 
1658.12  &  bl    & 7.51    & 7.50 &   +0.01     \\
4932.04  &  bl    & 7.75    & 7.74 &   +0.01     \\  
5052.14  &  6.1   & 7.69    & 7.68 &   +0.01    \\           
9088.51  &   79   & 7.83    & 7.88 &   -0.05    \\  
9111.80  &   92   & 7.80    & 7.86 &   -0.06    \\  
9405.73  &  107   & 7.49    & 7.54 &   -0.05    \\ 
9658.43  &   97   & 7.85    & 7.96 &   -0.11    \\\hline  
Mean C\ione\ &    & 7.71    & 7.74 &           \\ 
$\sigma$ &        & 0.15    & 0.15 &           \\\hline 
C\ii\     &         &      &     &           \\
1335.70  &  bl      & 7.64   & 7.64 &   0.00      \\  
1323.86\multirow{4}{*}{$\Bigg\}$}  & \multirow{4}{*}{bl}       &\multirow{4}{*}{7.74} &  \multirow{4}{*}{7.74} &  \multirow{4}{*}{0.00}   \\
1323.91&         &          &   &      \\
1323.95&         &          &   &      \\
1324.00&         &          &   &      \\\hline  
Mean C\ii&       &  7.69    & 7.69 &  \\
$\sigma$ &        & 0.15    & 0.15 &           \\\hline 
Mean C\ione+C\ii\ & & 7.71  & 7.73     &            \\
$\sigma$ &        &  0.14   & 0.14   &           \\ \hline 
  \end{tabular}
% \end{minipage}
\end{table}

\subsection{Determination of stellar carbon abundances}

\begin{table*}
 \begin{minipage}{350mm}
     \caption{NLTE abundances and abundance corrections for the program stars.  }
        \label{tab6} 
  \begin{tabular}{lccccccccccccccc}
  \hline 
  \hline 
 & \multicolumn{3}{c}{$\iota$~Her}  &  \multicolumn{3}{c}{$\pi$~Cet}     &  \multicolumn{3}{c}{HD~22136}       & \multicolumn{3}{c}{ 21~Peg}        & \multicolumn{3}{c}{HD~73666}      \\ 
  $\lambda$,\AA\ & {\tiny NLTE} &{\tiny LTE} & {\tiny$\Delta_{\rm NLTE}$}& {\tiny NLTE}& {\tiny LTE} & {\tiny$\Delta_{\rm NLTE}$}& {\tiny NLTE}& {\tiny LTE}& {\tiny$\Delta_{\rm NLTE}$}&{\tiny NLTE}& {\tiny LTE}& {\tiny$\Delta_{\rm NLTE}$}&{\tiny NLTE}& {\tiny LTE}   &{\tiny$\Delta_{\rm NLTE}$}\\\hline
C\ione\   &         &       &         &       &       &             &         &     &       &         &       &          &       &       &            \\
6007.17   &         &       &         &       &       &             &         &     &       &         &       &          &  8.61 & 8.63  & -0.02            \\
6012.22   &         &       &         &       &       &             &         &     &       &         &       &          &  8.60 & 8.62  & -0.02            \\
6013.21   &         &       &         &       &       &             &         &     &       &         &       &          &  8.46 & 8.48  & -0.02            \\
6014.83   &         &       &         &       &       &             &         &     &       &         &       &          &  8.59 & 8.61  & -0.02            \\
7111.46   &         &       &         &       &       &             &         &     &       &  8.31   & 8.18  & +0.13    &  8.48 & 8.54  & -0.06            \\
7113.17   &         &       &         & 8.57  & 8.14  &   +0.43     &         &     &       &  8.23   & 8.10  & +0.13    &  8.45 & 8.52  & -0.07            \\
7115.17\multirow{2}{*}{$\Big\}$} &         &       &      &       &       &                 &         &     &       &        \multirow{2}{*}{8.42} & \multirow{2}{*}{8.48} & \multirow{2}{*}{-0.06} &       &       &      \\
7115.18   &         &       &         &       &       &             &         &     &       &         &       &          &       &       &                  \\
7116.98   &         &       &         &       &       &             &         &     &       &  8.28   & 8.15  & +0.13    &  8.58 & 8.64  & -0.06            \\
7119.65   &         &       &         &       &       &             &         &     &       &  8.35   & 8.22  & +0.13    &  8.64 & 8.70  & -0.06            \\
4932.04   &         &       &         &       &       &             &         &     &       &  8.48   & 8.08  & +0.40    &       &       &                  \\
5039.06   &         &       &         &       &       &             &         &     &       &         &       &          &  8.52 & 8.50  & +0.02            \\
5052.14   &         &       &         &       &       &             &         &     &       &  8.26   & 7.81  & +0.45    &  8.53 & 8.51  & +0.02            \\
5380.32   &         &       &         &       &       &             &         &     &       &  8.22   & 7.66  & +0.56    &  8.53 & 8.51  & +0.02            \\
4762.52   &         &       &         &       &       &             &         &     &       &  8.46   & 8.40  & +0.06    &       &       &                  \\
4766.66   &         &       &         &       &       &             &         &     &       &  8.35   & 8.29  & +0.06    &       &       &                  \\
4770.02   &         &       &         &       &       &             &         &     &       &  8.49   & 8.43  & +0.06    &  8.47 & 8.50  & -0.03            \\
4771.73   &         &       &         &       &       &             &         &     &       &  8.46   & 8.40  & +0.06    &       &       &                  \\
4775.89   &         &       &         &       &       &             &         &     &       &  8.39   & 8.33  & +0.06    &  8.72 & 8.75  & -0.03            \\
8335.14   & 8.37    &       & $e$     & 8.49  &       &     $e$     &   8.44  &     &   $e$ &  8.22   &      & $e$       &  8.71 & 8.77  & -0.06            \\
9061.43   & 8.36    &       & $e$     &       &       &             &         &     &       &  8.35   & 8.48  & -0.13    &       &       &                  \\
9062.49   &         &       &         &       &       &             &         &     &       &  8.52   & 8.65  & -0.13    &       &       &                  \\
9078.28   & 8.51    &       & $e$     &       &       &             &         &     &       &  8.39   & 8.46  & -0.07    &       &       &                  \\
9088.51   & 8.34    &       & $e$     &       &       &             &  8.51   &8.01 &+0.50  &  8.40   & 8.49  & -0.09    &  8.58 & 8.90  & -0.32            \\
9111.80   &         &       &         &       &       &             &  8.42   &8.07 &+0.35  &  8.44   & 8.57  & -0.13    &  8.60 & 8.97  & -0.37            \\
9405.73   & 8.52    &       & $e$     & 8.64  &       &   $e$       &         &     &       &  8.50    &      &  $e$     &  8.55 & 8.61  & -0.06            \\
9603.02   &         &       &         &       &       &             &         &     &       &  8.35   & 8.40  & -0.05    &       &       &                  \\
9658.43   & 8.36    &       & $e$     &       &       &             &         &     &       &  8.40   & 8.54  & -0.14    &  8.61 & 9.02  & -0.41            \\\hline                                                                                                                              
Mean C\ione\ & 8.42 &       &         &8.57   & 8.14  &             & 8.46    & 8.04&       &  8.38   & 8.36  &          &  8.57 & 8.63  &           \\
$\sigma$  & 0.08    &       &         &0.08   & --    &             & 0.05    & 0.04&       &  0.09   & 0.26  &          &  0.08 & 0.17  &           \\\hline                                                                                                                              
C\ii\     &         &       &         &       &       &             &         &     &       &         &       &          &       &       &                  \\
3918.97   &  8.38   & 8.28  & +0.10   &  8.46 & 8.44  & +0.02       &  8.36   &8.35 &+0.01  &  8.39   & 8.38  & +0.01    &       &       &                  \\
3920.68   &  8.45   & 8.35  & +0.10   &       &       &             &         &     &       &         &       &          &       &       &                  \\
4267.00   &  8.33   & 8.26  & +0.07   &  8.44 & 8.43  & +0.01       &  8.41   &8.40 &+0.01  &  8.35   & 8.34  & +0.01    &  8.54 & 8.54  &0.00              \\
4267.26   &         &       &         &  8.44 & 8.43  & +0.01       &         &     &       &  8.36   & 8.35  & +0.01    &       &       &                  \\
6578.05   &  8.71   & 9.13  & -0.42   &  8.50 & 8.50  & 0.00        &  8.39   &8.44 &-0.05  &  8.35   & 8.36  & -0.01    &       &       &                  \\
6582.88   &  8.55   & 8.97  & -0.42   &  8.54 & 8.54  & 0.00        &  8.49   &8.54 &-0.05  &  8.41   & 8.42  & -0.01    &       &       &                  \\
5132.95   &  8.38   & 8.46  & -0.08   &       &       &             &         &     &       &         &       &          &       &       &                  \\ 
5133.28   &  8.37   & 8.45  & -0.08   &       &       &             &         &     &       &         &       &          &       &       &                  \\
5137.25   &  8.43   & 8.51  & -0.08   &       &       &             &         &     &       &         &       &          &       &       &                  \\
5139.17   &  8.35   & 8.43  & -0.08   &       &       &             &         &     &       &         &       &          &       &       &                  \\
5143.49   &  8.38   & 8.45  & -0.07   &       &       &             &         &     &       &         &       &          &       &       &                  \\
5145.16   &  8.41   & 8.48  & -0.07   &  8.37 & 8.40  & -0.03       &         &     &       &         &       &          &       &       &                  \\
5151.09   &  8.36   & 8.42  & -0.06   &       &       &             &         &     &       &         &       &          &       &       &                  \\
7231.33   &  8.29   & 8.18  & +0.11   &       &       &             &         &     &       &         &       &          &       &       &                  \\ 
7236.41   &         &       &         &  8.38 & 8.35  & +0.03       &         &     &       &         &       &          &       &       &                  \\
7237.17   &  8.41   & 8.34  & +0.07   &  8.37 & 8.34  & +0.03       &         &     &       &         &       &          &       &       &                  \\\hline                                                                                                                              
Mean C\ii\ &  8.43   & 8.58  &         &  8.44 &  8.45 &             & 8.42    &8.44 &       &  8.37   & 8.37  &          & 8.54  & 8.54  &           \\ 
$\sigma$  &  0.10   & 0.26  &         &  0.06 &  0.07 &             & 0.03    &0.08 &       &  0.03   & 0.03  &          & --    & --    &           \\\hline                                                                                                                              
Mean    &  8.43   & 8.58  &         &  8.45 &  8.41 &             & 8.43    &8.34 &       &  8.38   & 8.36  &          & 8.57  & 8.62  &           \\  
$\sigma$  &  0.10   & 0.26  &         &  0.09 &  0.11 &             & 0.05    &0.21 &       &  0.09   & 0.23  &          & 0.08  & 0.16  &           \\\hline                                                                                                                                
  \end{tabular}
  \begin{tablenotes}  
\item[a] {\bf Notes.} Emission lines are marked by symbol $e$. 
\end{tablenotes}  
 \end{minipage}
\end{table*}

In this section, we derive the carbon abundances of the selected stars from C\ione\ and C\ii\ lines using the atomic data from Table~\ref{tab1}. The results for individual stars are as follows.

{\bf Vega:} no emission line of C\ione\ is observed in the Vega's spectrum. For the lines longwards 10480\,\AA\ we adopted equivalent widths (EW) from \citet{1982ApJ...254..663L}. All the C\ione\ lines can be divided into two groups. Lines in the visible spectral range ($\lambda \le$ 7115.2\,\AA, 10 lines) are weak, 
with EW $<$ 50~m\AA, while lines in the near-IR spectral range (10\,123.9\,\AA\ $\le \lambda \le$ 10\,754\,\AA, 10 lines) are strong, with EW $>$ 100~m\AA. NLTE and LTE abundances from individual lines are presented in Table~\ref{tab4}. Everywhere, NLTE leads to strengthened lines, however, the effect is small, with $\Delta_{\rm NLTE} \le$ 0.07~dex by absolute value, for weak lines because they form in deep atmospheric layers. We note a large abundance discrepancy of 0.40~dex between the two weak lines, 4771 and 7113\,\AA, that cannot be removed by NLTE. A similar difference was also obtained by \citet{2001AA...379..936P} and \citet{1990AA...237..125S}. 

For strong lines, $\Delta_{\rm NLTE}$ varies mainly between $-0.28$ and $-0.64$~dex. Applying the treated model atom, we can reconcile the carbon abundance from the three strong (log~$\epsilon_{C}$ = 8.34$\pm$0.06) and 10 weak (log~$\epsilon_{\rm C}$ = 8.32$\pm$0.13) lines, while, in LTE, the IR lines give systematically higher abundances compared with the visible lines. 
Using EW from \citet{1982ApJ...254..663L} leads to a large line-to-line scatter of determined abundances, however, 
the mean NLTE abundance is consistent with that from
visible and near-IR lines, C\ione\ 9078, 9088, 9111\,\AA\ obtained in this study from
analysis of observed spectrum of Vega.  

In Table~\ref{tab4} we show also the NLTE abundances derived by \citet{2001AA...379..936P} and \citet{1990AA...237..125S}. 
It is worth noting that their abundances were reduced to the oscillator strengths of Table \ref{tab1}.
Our mean NLTE abundance derived from all C\ione\ lines, log~$\epsilon_{\rm C}$ = 8.34$\pm$0.13, is consistent with the error bars found in literature. We note, in particular, the C\ione\ 9078, 9088, 9111\,\AA\ lines, with large departures from LTE, for which the agreement is perfect.

We inspected an impact of varying collisional data on the derived carbon abundances of Vega.
In addition to our basic collisional recipe, WZB13+R94+vR62, Table\,\ref{tab4} presents the NLTE abundances and abundance corrections for R94+vR62 and vR62, as described above. The abundance difference between WZB13+R94+vR62 and R94+vR62 is small and nowhere exceeds 0.05~dex. However, applying rough theoretical approximations, i.e. the recipe vR62, leads to underestimated NLTE effects, in particular, for the strong lines. For example, the abundance difference between WZB13+R94+vR62 and vR62 amounts to 0.27~dex for C\ione\ 10691\,\AA.

{\bf Sirius:} the element abundances were derived from the C\ione\ and C\ii\ lines in a wide spectral region from UV to near-IR (Table~\ref{tab5}). The NLTE effects were found to be overall small. For lines of C\ii\ and lines of C\ione\ in the UV and visible range, $\Delta_{\rm NLTE} \le$ 0.01~dex. This can be easily understood. C\ii\ is the dominant ionisation stage. The C\ione\ UV lines arise in the transitions between the lowest levels, which are closely coupled together (see the departure coefficients in Fig.~\ref{DC}). Both C\ione\ visible lines are weak. 
For C\ione\ near-IR lines, the NLTE corrections are negative and do not exceed 0.11~dex in absolute value. 
The mean NLTE abundances from lines of the two ionization stages, C\ione\ and C\ii, appear to be consistent. Although, we note rather large line-to-line scatter for C\ione. 

Small influence of NLTE on the carbon abundance determination for Sirius makes possible to compare our results with the data obtained by different studies under the LTE assumption.
A complete analysis of the literature data was given by \citet{2011AA...528A.132L} who analysed the UV spectrum of Sirius. The average abundance, 
log~$\epsilon_{\rm C}$ = 7.79$\pm$0.18, deduced by \citet{2011AA...528A.132L} (his Table~1) agrees well with our  determination. 

For the other five stars of our programme, the abundance results are presented in Table~\ref{tab6}. In all stars but the coolest one, HD~73666, few/all C\ione\ lines are observed in emission.
These lines are marked by '$e$' in the corresponding $\Delta_{\rm NLTE}$ columns. For comparison, we adopted the solar carbon NLTE abundance, log~$\epsilon_{\rm C}$ = 8.43, derived by \citet{2015MNRAS.453.1619A} from the solar C\ione\ atomic lines. 

{\bf HD~73666:} there are plenty of C\ione\ lines in the spectrum of this Blue Straggler,  while C\ii\ is represented by the only weak line.
The mean NLTE abundances from lines of the two ionization stages, C\ione\ and C\ii, were found to be consistent, and they are 0.1~dex higher than the solar value, in line with the 
overall metallicity of HD~73666 \citep{2007A&A...476..911F}. 

{\bf 21~Peg:} this is the coolest star in our sample that shows an emission in the C\ione\ near-IR lines at 8335\,\AA\ and 9405\,\AA.
As discussed in Sect.\,\ref{Sect:departure}, the NLTE abundance corrections are positive and large for the C\ione\ visible lines, with $\Delta_{\rm NLTE}$ up to +0.56~dex. 
This explains low carbon abundance obtained from these lines by \citet{2009A&A...503..945F} under the LTE assumption. We determined the element abundance from the C\ione\ emission lines, too, and they appear to be consistent with that from the C\ione\ absorption lines. The departures from LTE are minor for C\ii. NLTE provides consistent abundances from lines of the two ionisation stages, C\ione and C\ii, including the C\ione emission lines, in contrast to LTE, where an abundance difference between the individual lines reaches 0.5~dex and the emission lines cannot be reproduced.
The best NLTE fits of the C\ione\ absorption and emission lines in 21~Peg are shown in Fig.\,\ref{pics}. 
The obtained carbon NLTE abundance of 21~Peg is close to the solar value.

{\bf HD~22136:} we could measure only three lines of C\ione\ and four lines of C\ii\ in this star because 
of heavy blending the near-IR spectrum by the telluric lines. Element abundance derived from the C\ione\ 9405\,\AA\ emission line is consistent with the NLTE abundances from the C\ione\ and C\ii\ absorption lines. The obtained carbon NLTE abundance of HD~22136 is very similar to the solar one. 

{\bf $\pi$~Cet:} this star is hotter by 100~K than HD~22136 and slightly faster rotating.
Two emission lines, C\ione\ 8335 and 9405\,\AA, were detected in its spectrum. 
The only measurable absorption line of C\ione\ is 7113\,\AA. 
NLTE abundance corrections for the C\ii\ lines are mainly positive and, for each line, $\Delta_{\rm NLTE}$ does not exceed 0.03~dex in absolute value.
Carbon abundances from the absorption and emission lines of C\ione\ agree with each other, but an average abundance from the C\ione\ lines is higher than that derived from the C\ii\ lines, although a difference of 0.13~dex is still within 2$\sigma$.  

{\bf $\iota$~Her:} this is the hottest star of our sample. No C\ione\ line was detected in the visible range. 
All the C\ione\ near-IR lines appear in emission.
NLTE abundance corrections have different sign for different lines of C\ii.
For most lines they do not exceed 0.11~dex in absolute value, but they are large and negative, with $\Delta_{\rm NLTE} = -0.42$~dex, for strong C\ii\ 6578, 6582~\AA\ lines.
Our NLTE analysis provides consistent abundances from different groups of lines, i.e. the C\ione\ emission  lines and C\ii\ absorption lines. The best NLTE fits of the C\ione\ emission lines and C\ii\ absorption lines are shown in Fig.\,\ref{pics}. 
The obtained carbon abundance of $\iota$~Her, log~$\epsilon_{\rm C}$ = 8.43$\pm$0.10, is essentially solar. Similar result, log~$\epsilon_{\rm C}$ = 8.40$\pm$0.07, was obtained by \citet{2012A&A...539A.143N} in their NLTE analysis of the C\ii\ lines in $\iota$~Her. 
Larger dispersion in our abundance determination is mainly caused by the two lines, C\ii~6578~\AA\ and 6582~\AA, which give about 0.2~dex higher abundance compared with that from the remaining lines, while no significant abundance discrepancy between different lines was obtained by \citet{2012A&A...539A.143N}.

\section{Conclusions}\label{Sect:Conclusions}

The motivation of this work was to solve two problems in stellar astrophysics: the search for 
explanation of the appearance of emission lines in C\ione\ 
in the near-IR spectral region and 
a reliable determination of carbon abundances for AB-type stars.
We constructed a comprehensive model atom for C\ione\ -- C\ii\ using the most up-to-date atomic data and evaluated the NLTE
line formation for C\ione\ and C\ii\ in classical 1D models representing the atmospheres of A and late B-type stars.

Our NLTE calculations predict that some lines of C\ione\ in the near IR spectral range 
may appear as emission lines depending on the atmospheric parameters. The emission appears first in the C\ione\ 8335\,\AA\ and 9405\,\AA\ singlet lines at effective temperature of 9250~K to 10\,500~K depending
on the value of log~$g$. It is strengthened toward higher \Teff, reaches a maximal level at \Teff\ = 16\,000~K (log~$g$ = 3) and almost disappears at \Teff\ = 22\,000~K. The C\ione\ triplet lines at
 9061--9111\,\AA\ and 9603--9658\,\AA\ come into emission at \Teff\ $>$ 15\,000~K (log~$g$ = 4). The mechanisms driving the C\ione\ emission can be understood as follows.
A prerequisite of the emission phenomenon is the overionization-recombination mechanism resulting in a depopulation of the lower levels of C\ione\ to a greater extent than the upper levels.
Extra depopulation of 3s$^{1}$P$^{\circ}$ and 3s$^{3}$P$^{\circ}$, which are the lower levels of the transitions corresponding to the listed near-IR lines, 
can be caused by photon loss in the UV lines C\ione\ 2479, 1930, and 1657\,\AA. In the models with \Teff\ of about 10\,000~K only, C\ione\ 2479\,\AA\ plays a role draining population of the 3s$^{1}$P$^{\circ}$ singlet level effectively in the layers, where the C\ione\ 8335 and 9405\,\AA\ lines form, while these layers are optically thick for radiation at 1930 and 1657\,\AA. 
With increasing \Teff, formation depths of all the C\ione\ UV lines shift inwards, resulting in a 
depopulation of not only 3s$^{1}$P$^{\circ}$, but also 3s$^{3}$P$^{\circ}$ and the lower levels of the emission triplet lines at 9061 -- 9111\,\AA\ and 9603 -- 9658\,\AA.

 Our theoretical results were confirmed with observations of the reference stars.
The stellar sample consists of seven bright and
apparently slow-rotating A- and late B-type stars in the solar neighbourhood. Our analysis is based on  
high S/N and high-resolution spectra with a broad wavelength coverage.

The C\ione\ emission lines were measured in the four hottest stars, with \Teff\ $\ge$ 10\,400~K, and they were well reproduced in our NLTE calculations.
For each star, the mean NLTE abundances from lines of the two ionisation stages, C\ione\ and C\ii, including the C\ione\ emission lines, were found to be consistent. 
Thus, we settled the dispute on whether the C\ione\ emission in the late-B stars is produced by the circumstellar disc, or 
vertical stratification / horizontal inhomogeneity of the atmosphere, or the NLTE effects in the atmosphere.

The six of our stars reveal highly uniform and close-to-solar carbon abundance. We confirm a significant underabundance of carbon in Sirius, with [C/H] = $-0.72$.

We show an importance of applying accurate atomic data to the statistical equilibrium calculations.
In particular, the C\ione\ emission phenomenon turns out to be extremely sensitive to varying electron-impact excitation rates. If the latter are not accounted for properly, the stellar C\ione\ emission lines cannot be reproduced. The results obtained in this study favour the collisional data for C\ione\ from predictions of \citet{2013PhRvA..87a2704W}.

It is worth noting, WELs of Mg\ii, Si\ii, P\ii, Ca\ii, Cr\ii, Fe\ii, Ni\ii, Cu\ii, and Hg\ii\ were detected in the near-UV and visible spectral regions for several stars not showing any trace of chromosphere \citep{2000ApJ...530L..89S, 2000AA...362L..13W, 2004AA...418.1073W,2007AA...475.1041C}. 
According to \citet{2008CoSka..38..279W}, WELs of some metals are observed in sharp-lined spectra of mid- to late-B type stars. The WELs are detected over a range of element abundance and are found among both chemically-normal and chemically-peculiar stars. The NLTE line--formation calculations for these specific ions are highly desirable to understand mechanisms of the observed emission.

{\it Acknowledgements.}
This research is based on observations obtained with
MegaPrime/MegaCam, a joint project of CFHT and CEA/IRFU,
at the Canada--France--Hawaii Telescope (CFHT) which is operated by the National Research Council (NRC) of Canada, the Institut National des Science de l$'$Univers of the Centre National de la
Recherche Scientifique (CNRS) of France, and the University of
Hawaii. 
We thank Oleg Zatsarinny for providing us the data on effective collision strengths for the C\ione\ transitions published in \citet{2013PhRvA..87a2704W}.
This work was supported in part by the Russian Foundation for Basic Research (grants 16-32-00695 and 15-02-06046).

%%%%%%%%%%%%%%%%%%%%%%%%%%%%%%%%%%%%

% \bsp % ``This paper has been produced using the ...''

%\label{lastpage}

\bibliography{carbon}
\bibliographystyle{mn2e}

\end{document}